\documentclass[aps,prd,showpacs,preprintnumbers
,epsf,nofootinbib]{revtex4}
\usepackage{amsfonts,latexsym,eucal,amsmath,amssymb,times}
\usepackage[dvips]{graphicx}
\usepackage[usenames]{color}

\newcommand{\ie}{{i.e.}}
\newcommand{\eg}{{\it e.g.}}
\newcommand{\etal}{{et al.}}

\newcommand{\wlm}{{\omega lm}}

\newcommand{\mwlm}{{-\omega\,l\,-m}}

\newcommand{\Dl}{{\Delta\lambda}}

\definecolor{red  }{rgb}{1,0,0}
\definecolor{blue }{rgb}{0,0,1}
\definecolor{green}{rgb}{0,1,0}

\begin{document}
\draft

\thispagestyle{empty}

\title{Evolution of the Carter constant for resonant inspirals 
into a Kerr black hole: I. The scalar case}
\date{\today}
\author{Soichiro Isoyama$^{1}$}
\email{isoyama_at_yukawa.kyoto-u.ac.jp}
\author{Ryuichi Fujita$^{2}$}
\email{ryuichi.fujita_at_uib.es}
\author{Hiroyuki Nakano$^{1,3}$}
\email{hinakano_at_yukawa.kyoto-u.ac.jp}
\author{Norichika Sago$^{4}$}
\email{sago_at_artsci.kyushu-u.ac.jp}
\author{Takahiro Tanaka$^{1}$}
\email{tanaka_at_yukawa.kyoto-u.ac.jp}
\affiliation{\,\\ \,\\
$^{1}$ Yukawa Institute for Theoretical Physics, Kyoto university,
Kyoto, 606-8502, Japan \\
$^{2}$ Departament de F\'isica, Universitat de les Illes Balears, 
Palma de Mallorca, E-07122 Spain \\
$^{3}$ Center for Computational Relativity and Gravitation,
Rochester Institute of Technology,
New York 14623, USA \\
$^{4}$ Faculty of Arts and Science, Kyushu University,
Fukuoka 819-0395, Japan \\
}

\preprint{YITP-12-102}


\begin{abstract}
We discuss the inspiral of a small body around a Kerr black hole.
When the time scale of the radiation reaction is sufficiently 
longer than its orbital period, the leading-order orbital evolution
is described only by the knowledge of 
the averaged evolution of constants of motion, 
{\ie}, the energy, azimuthal angular momentum and the Carter constant.  
Although there is no conserved current composed of the perturbation field 
corresponding to the Carter constant, 
it has been shown that the averaged rate of change of the Carter constant 
can be given by a simple formula, when there exists a simultaneous 
turning point of the radial and polar oscillations. 
However, an inspiraling orbit may cross a resonance point, 
where the frequencies of the radial and polar orbital 
oscillations are in a rational ratio. 
At the resonance point, one cannot find a 
simultaneous turning point in general. 
Hence, even for the averaged rate of change of the Carter constant,  
a direct computation of the self-force, which is still challenging 
especially in the case of the Kerr background, seems to be necessary.
In this paper, we present a method of
computing  the averaged rate of change of the Carter constant  
in a relatively simple manner at the resonance point.
\end{abstract}

\maketitle


\section{Introduction}
\label{sec:Intro}
%
%
The study of the two-body problem in general relativity has attracted 
much attention in recent years in order to obtain a better understanding 
of the gravitational wave emission from a binary system. If one body is
much less massive than the other,
{i.e.}, when we consider extreme mass ratio 
inspirals of a compact object 
into a super massive black hole, 
the compact object can be treated as a point particle and the
two-body problem is handled within the self-force program
(see Refs. \cite{Barack:2009ux,Poisson:2011nh} and references therein).

When a nonspinning particle with mass $\mu$ is moving along a bound orbit 
around a Kerr black hole with mass {$M (\gg \mu$)} 
and the spin parameter $a$, the long-term evolution of the orbit
due to the  radiation reaction of the gravitational wave emission is
dictated by the long-time averaged rates of change of three constants
of motion: the energy $E$, azimuthal angular momentum $L$ and
Carter constant $Q$ \cite{Carter:1968rr}
since the characteristic time scale of
the secular orbital evolution is
sufficiently longer than the orbital period.
The orbital phase errors originating from the other effects scale as 
$O((M/\mu)^0)$.
This particular approximation of the orbital evolution is referred to 
as the adiabatic approximation 
\cite{Mino:2005an,Tanaka:2005ue,Pound:2007th,Pound:2007ti,Hinderer:2008dm}.

On average, the energy and azimuthal angular momentum losses of the particle 
are balanced with those carried by gravitational waves 
to the infinity or into the Kerr horizon 
due to their global conservation laws.
This is nothing but the well established balance argument.
Although we do not have any balance argument for the Carter constant, 
based on Mino's argument~\cite{Mino:2003yg},
Sago {et al.}~\cite{Sago:2005gd,Sago:2005fn} 
(see also Drasco~{\etal}~\cite{Drasco:2005is} for the scalar case)
have succeeded in deriving a formula for
the averaged rate of change of the Carter constant 
$\langle dQ/dt \rangle$
similar to that obtained by the balance argument for
the energy and azimuthal angular momentum.

However, the above-mentioned work 
assumes that the particle's radial and polar orbital frequencies,
$\Omega_r$ and $\Omega_{\theta}$, are in an irrational ratio.
When $\Omega_r$ and $\Omega_{\theta}$ cross a resonance point,
at which $\Omega_r / \Omega_{\theta} = j_r/ j_{\theta}$ 
holds with coprime integers $j_r$ and $j_{\theta}$,
the above formula is not correct, and the
phase error due to ignorance of the correct formula
can be as large as $O((M/\mu)^{1/2})$.
This was first pointed out by Mino~\cite{Mino:2005an}
(see also Ref.~\cite{Tanaka:2005ue}).
Recently, Flanagan and Hinderer~\cite{Flanagan:2010cd} investigated
this issue quantitatively within the post-Newtonian approximation.
For the inspiral in the extreme mass ratio regime $M \gg \mu$, 
this phase error is much larger than that from 
the first-order conservative self-force or the 
long-time averaged second-order dissipative self-force,
which scales as $O((M/\mu)^0)$
\cite{Tanaka:2005ue, Hinderer:2008dm, Pound:2007th, Huerta:2008gb,
Yunes:2010zj, Isoyama:2012bx}.
For this reason, it is very desirable to derive a formula for
$\langle dQ/dt \rangle$ in the adiabatic
approximation for resonance orbits.

The main differences from the off-resonance case in the derivation of
$\langle dQ/dt \rangle$ are the following two points.
First, the orbit is no longer characterized by the three constants of 
motion alone. 
Instead, as was pointed out by Flanagan~{\etal}~\cite{Flanagan:2012kg}, 
the orbit depends on the relative offset phase specified by $\Dl$, 
the separation between the instances when the $r$- and 
$\theta$-oscillations reach their minima.
The offset phase also slowly evolves compared to the orbital period
near the resonance point. 
Thus, the resonance orbit must be characterized 
by four parameters: $\{E, L_z, Q,  \Delta \lambda \}$.
Second, we cannot use Mino's argument~\cite{Mino:2005an}, 
which proves that $\langle dQ/dt \rangle$ can be
evaluated by using the radiative field defined by 
``the half-retarded minus half-advanced field'', which is 
by definition regular at the location of the particle.
The key ingredient of Mino's argument is an 
invariance of the geodesic under the transformation 
$(t,r,\theta,\phi) \to (-t,r,\theta,-\phi)$ that is 
assured if and only if the $r$- and $\theta$-components of the body's four 
velocity vanish simultaneously at a certain time, {\it i.e.},
when we can choose $\Dl = 0$.
Clearly, this condition is not satisfied for resonance geodesics.

Our final goal is to obtain the evolution of resonant inspirals 
with a phase accuracy better than $O((M/\mu)^0)$.
Toward this ambitious goal, we here begin with deriving  
a practical formula of $\langle dQ/dt \rangle$
for the resonant inspirals.
In the resonance case, we find that $\langle dQ/dt \rangle$
acquires contributions from both the radiative and symmetric fields, 
the latter defined by 
``the half-retarded {\it plus} half-advanced field''.
In general, the symmetric field diverges at the particle's location, 
thus requires some regularization.
In a later section, we will propose a new regularization 
scheme for the symmetric part of $\langle dQ/dt \rangle$.
The key point of our regularization method is that it is compatible
with the Teukolsky formalism~\cite{Teukolsky:1973ha}. 
This makes the calculation much simpler than the self-force calculation, 
which has already proved challenging numerically for a particle orbiting 
around a Schwarzschild black hole
(see Ref.~\cite{Barack:2009ux} for recent progress).

In this paper, to avoid complications,
we focus on a point particle with mass $\mu$ 
coupled to a massless scalar field with the charge $q$.
Although in principle there seems to be no essential obstacle to extending 
our scalar formalism to the gravitational case, 
the latter is more involved since it requires the reconstruction of
metric perturbations, which is a little complicated in the Kerr case 
\cite{Whiting:2005hr,Shah:2012gu}.

The remainder of this manuscript is organized as follows.
In Sec.~\ref{sec:Scalar-model}, 
we begin with describing the model of a point scalar particle 
coupled to a massless scalar field. 
We review the generic bounded geodesic motion 
around a Kerr black hole, particularly focusing on the resonance case.
The retarded solution of the scalar field equation expressed
by the mode decomposition is also summarized there.
We derive a general formula for
the long-time averaged rate of change 
of the Carter constant in Sec.~\ref{sec:dIdt}. 
We separate the formula into two parts, 
the radiative and symmetric parts 
that originate from the radiative and symmetric
fields, respectively. The result for the radiative part is shown
in Sec.~\ref{subsec:dQdt-R}, and that for the symmetric
part is presented in Sec.~\ref{subsec:dQdt-Sym}, where we also
discuss the mode sum regularization of the symmetric part 
since it diverges at the particle's location.
Finally, we summarize this paper in Sec.~\ref{sec:Conclusion}.
To clarify the impact of our work, we also show how 
the long-term evolution of the resonance orbit is described in
the adiabatic approximation.
In Appendix \ref{app:evolution}, we discuss a possibility that
the resonance effect on the phase error might be larger than
$O((\mu/M)^{-1/2})$.
Some detailed derivations of formulas
related to the regularized symmetric
part are summarized in Appendix \ref{app:simplify-dQdt-sym}.

Throughout this paper, we use geometrical units $G = c = 1$ and
the sign convention of the metric is $(-,+,+,+)$.
We basically follow the notation used by
Drasco~\etal~\cite{Drasco:2005is}.

\section{Preliminaries}
\label{sec:Scalar-model}
%
%
We consider a point scalar particle with the bare rest mass $\mu=1$
\footnote{
Strictly speaking, the mass $\mu$ is not constant because of the self-force
effect. However, the correction due to this time dependence
is $O((M/\mu)^0)$ in the orbital phase, {\it i.e.}, suppressed as well
as the conservative self-force correction. Since we focus on the
correction of $O((M/\mu)^{1/2})$, we neglect the change in the mass
in this paper.}
and scalar charge $q$, moving in the Kerr spacetime.
The Kerr metric in the Boyer--Lindquist coordinates
is given by
\begin{equation}\label{Kerr}
g_{\mu \nu} dx^{\mu} dx^{\nu} = 
- \left( 1 - \frac{2 M r}{\Sigma} \right) dt^2 
- \frac{4 M a r {\rm{sin}^2 \theta} }{\Sigma} dt d \phi 
+ \frac{\Sigma}{\Delta} dr^2 + \Sigma d \theta^2 
+ \left( r^2 + a^2 + \frac{2 M a^2 r}{\Sigma} \sin^2 \theta \right)
\sin^2 \theta \, d \phi^2 ,
\end{equation}
where $M$ and $a$ are the mass and the Kerr parameter, respectively, and
\begin{equation}\label{Kerr-hojo}
\Sigma := r^2 + a^2 \cos^2 \theta, 
\qquad
\Delta := r^2 - 2 M r + a^2. 
\end{equation}
There are two Killing vectors associated with the
stationarity and axisymmetry of the Kerr spacetime:
\[
 \xi_{(t)}^{\mu} := (\partial_t)^{\mu} = (1, 0, 0, 0), \quad
 \xi_{(\phi)}^{\mu} := (\partial_{\phi})^{\mu} = (0, 0, 0, 1).
\]
The Kerr spacetime also has an irreducible Killing tensor
given by
\begin{equation}\label{CKY}
K_{\mu \nu}
:= 2 \Sigma \ell_{( \mu} n_{\nu)} + r^2 g_{\mu \nu} 
= 2 \Sigma m_{( \mu} \bar{m}_{\nu)} -a^2 \cos^2 \theta g_{\mu \nu},
\end{equation}
with the Kinnersley null tetrad 
defined by $\ell^{\mu} := (r^2 + a^2,\Delta,0,a) / \Delta$, 
$n^{\mu} := (r^2 + a^2,-\Delta,0,a)/(2 \Sigma)$ 
and 
$m^{\mu} := (i a \sin \theta,0,1,i/\sin \theta) 
/ (\sqrt{2} (r + i a \cos \theta))$, 
where $\ell^{\mu} n_{\mu} = -1$ and $m^{\mu} \bar{m}_{\mu} = 1$ 
while all the other inner products vanish.

\subsection{Bounded geodesic motion around a Kerr black hole in resonance}
\label{subsec:Geodesic}
%
Neglecting the effect of self-interaction, a particle
can be regarded as a test particle that
moves along a geodesic on the background spacetime,
$z^\mu (\tau) := 
(t(\tau), r(\tau), \theta(\tau), \phi(\tau))$ parametrized by
the proper time $\tau$. There are three integrals of motion 
from the symmetries of the Kerr spacetime:
the energy $E$, azimuthal angular momentum $L$ and 
Carter constant $Q$~\cite{Carter:1968rr}, 
respectively. 
These are expressed as
\begin{eqnarray}\label{constants}
{E}  &:=& 
-u^{\alpha} \xi_{\alpha}^{(t)} = 
\left( 1 - \frac{2 M r}{\Sigma} \right) u^{t}
+ \frac{2 M a r \sin^2 \theta}{\Sigma} u^{\phi},\cr
{L} &:=& u^{\alpha} \xi_{\alpha}^{(\phi)} = 
- \frac{2 M a r \sin^2 \theta}{\Sigma} u^{t} 
+\frac{(r^2 + a^2) - \Delta a^2 \sin^2 \theta}{\Sigma} \sin^2 \theta 
\, u^{\phi}, \cr
\label{Carter-const}
{Q} &:=& u^{\alpha} u^{\beta} K_{\alpha \beta} = 
\left({L \over \sin\theta} 
- a \sin \theta  E\right)^2 + a^2 \cos^2 {\theta} 
+ \Sigma^2 { (u^{\theta} )^2},
\end{eqnarray}
where $u^{\alpha} := d z^{\alpha} /d \tau$ is the particle's four
velocity 
(note that $E$ and $L$ are $O(\mu)$, and $Q$ is $O(\mu^2)$
if we recover $\mu$).

For the Kerr geodesics, it is convenient to parametrize the
particle's orbit as
$z (\lambda) := 
(t(\lambda), r(\lambda), \theta(\lambda), \phi(\lambda))$ 
with the ``Mino time'' $\lambda$, which is related to $\tau$ as
$d \lambda := d \tau / \Sigma $ \cite{Mino:2005an}. 
Using $\lambda$, the equations of motion are written as
\begin{eqnarray}
\label{geodesic-p}
&&\left( \frac{dr}{d \lambda} \right)^2 = R(r),
\qquad\quad
\left( \frac{d \cos \theta }{d \lambda} \right)^2 =\Theta (\cos
\theta)~, 
\label{geodesic-np}
\\
&&\frac{dt}{d \lambda}= V_{t \theta}(\theta) + V_{t r}(r)~,
\qquad
\frac{d \phi}{d \lambda} 
= V_{\phi \theta}(\theta) + V_{\phi r}(r),
\end{eqnarray}
where
\begin{eqnarray*}
R(r) &:=& P(r)^2 - \Delta (r^2 + Q), \\
P(r) &:=& E(r^2 + a^2) - a L, \\
\Theta (\cos \theta) &:=&
C - \{C + a^2 (1 - E^2)+ L^2 \}\cos^2 \theta 
+ a^2 (1 - E^2) \cos^4 \theta, \\
V_{t \theta}(\theta) &:=& - a ( a E \sin^2 \theta - L), 
\qquad
V_{t r}(r):= \frac{r^2 + a^2}{\Delta} P(r),\\
V_{\phi \theta}(\theta) &:=& -a E + \frac{L}{\sin^2 \theta}, 
\qquad \qquad 
V_{\phi r}(r):=\frac{a}{\Delta} P(r),
\end{eqnarray*}
with $C := Q - (a E - L)^2$.
Here it should be noted that the equations of $r$ and $\theta$ are
completely decoupled, and those of $t$ and $\phi$ are divided
into $r$ and $\theta$ -dependent parts.

First, the $r$- and $\theta$ -oscillations are independently periodic 
for a bound orbit:
\begin{equation} \label{eq:r-theta-motions}
r(\lambda)  = r(\lambda + n \Lambda_r),
\qquad
\theta(\lambda) = \theta (\lambda + k \Lambda_{\theta}),
\end{equation}
from Eqs~\eqref{geodesic-p}.
Here, $n$ and $k$ are integers and the periods $\Lambda_r$ and 
$\Lambda_{\theta}$ with respect to the Mino time $\lambda$ are 
given by 
\begin{equation}\label{Mino-period}
\Lambda_r := 2 \int_{r_{\rm min}}^{r_{\rm max}} \frac{dr}{\sqrt{R(r)}}~,
\qquad
\Lambda_{\theta} := 4 \int_{0}^{\cos \theta_{\rm min}} 
\frac{d (\cos \theta )}{\sqrt{\Theta (\theta)}}
\end{equation}
with the turning points $r_{\rm min},~r_{\rm max}$ and 
$\theta_{\rm min} (< \pi/2)$. For later convenience, we
introduce $\bar r$ and $\bar\theta$ which 
denote the fiducial solutions of Eqs.~\eqref{geodesic-np}
such that $\bar r(n\Lambda_r)=r_{\rm{min}}$ and 
$\bar \theta(k\Lambda_r)=\theta_{\rm{min}}$. 

The precise meaning of the resonance is that the orbital frequencies 
\[
\Upsilon_{r} := \frac{2\pi}{\Lambda_{r}}, \quad
\Upsilon_{\theta} := \frac{2\pi}{\Lambda_{\theta}}
\]
are related to each other as 
\begin{equation}
\frac{\Upsilon_{r}}{j_r} = 
\frac{\Upsilon_{\theta}}{j_{\theta}},
\end{equation}
where $j_r$ and $j_{\theta}$ are coprime integers.
We define the resonance frequency, $\Upsilon$, and
the corresponding period, $\Lambda$, as
\[
\Upsilon :=
\frac{\Upsilon_{r}}{j_r} = 
\frac{\Upsilon_{\theta}}{j_{\theta}}, \quad
\Lambda := \frac{2\pi}{\Upsilon}.
\]
In the resonance case, there is a difference between 
the Mino times of reaching the minima of the $r$- and $\theta$ -oscillations 
in general.
We call this difference the ``offset phase'' and denote it by $\Dl$. 
We note that $\Dl$ can be shifted by
\[
\Lambda' := \frac{\Lambda}{j_r j_\theta},
\]
depending on which minima we focus on. 
Therefore, we choose $\Dl$ such that $|\Dl|\leq \Lambda'/2$. 
Setting the origin of the $\lambda$ coordinate to satisfy 
$\theta(\lambda)=\bar \theta(\lambda)$, the orbit is 
written as 
\begin{equation}\label{r-motion}
r(\lambda)=\bar r(\lambda-\Dl)~,
\qquad
\theta(\lambda)=\bar \theta(\lambda)
\end{equation}
without loss of generality. 

Second, the motion in the $t$-direction is also separated into 
$r$- and $\theta$-dependent parts. We introduce the 
averaged values of $V_{tr}(r)$ and $V_{t \theta}(\theta)$ defined by 
\begin{eqnarray}
\label{Vt-averaged}
\langle V_{tr} \rangle := \frac{1}{\Lambda_r} 
\int_0^{\Lambda_r} d \lambda'\, V_{tr}
{[{\bar r}(\lambda')]},
\qquad
\langle V_{t \theta} \rangle := 
\frac{1}{\Lambda_{\theta}} \int_0^{\Lambda_{\theta}} 
d \lambda'\, V_{t \theta}
{[{\bar \theta} (\lambda')]}.
\end{eqnarray}
In integrating the $t$-component of the equation of motion,
the choice of the origin of the $t$-coordinate can be changed freely
because of the time-translation symmetry of the Kerr spacetime.
Then, without loss of generality,
$t(\lambda)$ is given by \cite{Drasco:2005is} 
\begin{eqnarray}
\label{geodesic-t}
t(\lambda) = \Upsilon_t \lambda 
+ \Delta t_r (\lambda-\Dl)
+ \Delta t_{\theta}(\lambda),
\end{eqnarray}
where 
$\Upsilon_t :=  \langle V_{t r} \rangle + \langle V_{t \theta} \rangle$, 
and 
\begin{eqnarray}\label{t-oscillation}
\Delta t_r (\lambda) := \int_0^{\lambda } d \lambda' 
\{ V_{tr}[\bar r(\lambda')] - \langle V_{tr} \rangle \},
\qquad
\Delta t_{\theta} (\lambda) := 
\int_0^{\lambda} d \lambda' 
\{ V_{t \theta}[\bar \theta (\lambda')] 
- \langle V_{t \theta} \rangle \}. 
\end{eqnarray}
The function $\phi(\lambda)$ is deduced in the same way as $t(\lambda)$.
Again considering the axial symmetry of the Kerr spacetime,
we have 
\begin{eqnarray}\label{phi-oscillation}
\phi(\lambda) &=& \Upsilon_{\phi} \lambda 
+ \Delta \phi_r (\lambda-\Dl)
+ \Delta \phi_{\theta} (\lambda)~,
\end{eqnarray}
where we define 
$\Upsilon_{\phi} :=  \langle V_{\phi r} \rangle + \langle V_{\phi \theta} 
\rangle$ and 
\begin{eqnarray}
\Delta \phi_r (\lambda) := \int_0^{\lambda } d \lambda' 
\{ V_{\phi r}[\bar r(\lambda')] - \langle V_{\phi r} \rangle \}~,
\qquad
\Delta \phi_{\theta} (\lambda) := 
\int_0^{\lambda} d \lambda' 
\{ V_{\phi \theta}[\bar \theta (\lambda')] 
- \langle V_{\phi \theta} \rangle \}
\end{eqnarray}
with
\begin{eqnarray}
\label{Vphi-averaged}
\langle V_{\phi r} \rangle := \frac{1}{\Lambda_r} 
\int_0^{\Lambda_r} d \lambda'\, V_{\phi r}
{[{\bar r}(\lambda')]},
\qquad
\langle V_{\phi \theta} \rangle := 
\frac{1}{\Lambda_{\theta}} \int_0^{\Lambda_{\theta}} 
d \lambda'\, V_{\phi \theta}
{[{\bar \theta} (\lambda')]}. 
\end{eqnarray}

From the above discussion, the orbit is expressed as 
\begin{equation}\label{z-lambda}
z(\lambda) = \bar z(\lambda - \Delta \lambda, \lambda)
\end{equation}
with the two-variable function 
\begin{equation}
\label{subbarz}
\bar z(\lambda_r,\lambda_\theta)
=[\Upsilon_t\lambda_\theta,0,0,\Upsilon_\phi\lambda_\theta]
+\Delta \bar z(\lambda_r,\lambda_\theta), 
\end{equation}
where the oscillating part of the orbit is denoted by 
\begin{equation}
\label{delta-barz}
\Delta \bar z(\lambda_r,\lambda_\theta)
= [\Delta t_r(\lambda_r) + \Delta t_\theta(\lambda_\theta), 
\bar r(\lambda_r), \bar \theta(\lambda_\theta), 
\Delta \phi_r(\lambda_r) + \Delta \phi_\theta(\lambda_\theta)]. 
\end{equation}

\subsection{The retarded solution of the scalar field equation}
\label{subsec:Scalar-field}
The equation of the scalar field induced by a charged
particle is given by
\begin{equation}\label{scalar-EOM}
g^{\mu\nu} \nabla_\mu \nabla_\nu \Phi (x)
= {\cal T}(x),
\qquad 
{\cal T}(x) :=
-{q} \int^{+\infty}_{-\infty} d \tau 
\frac{ \delta^{(4)} (x-z(\tau)) } { \sqrt{-g(z(\tau))} },
\end{equation}
where $\nabla_\mu$ and $g$ are the covariant differentiation
and the determinant for the background Kerr metric, $g_{\mu\nu}$,
in Eq.~(\ref{Kerr}), respectively 
(see Ref.~\cite{Drasco:2005is} for further details).
Assuming that there is no incoming scalar wave, we take
the retarded solution, $\Phi^{\rm (ret)}$.
To obtain the solution, we here construct the
retarded Green function in terms of mode functions.

From the fact that the scalar field equation is separable
in the Kerr spacetime~\cite{Teukolsky:1973ha, Brill:1972xj},
we can write the mode functions in the form of
\begin{equation}\label{complete-mode}
\pi_{\wlm}^{\flat}(t,r,\theta,\phi) := \frac{2}{\sqrt{r^2 + a^2}} 
e^{-i \omega t }
u_{\wlm}^{\flat}(r_{\ast})
S_{\wlm} (\theta, \phi),
\end{equation}
where $r_{\ast}$ is the tortoise coordinate defined by
$d r_{\ast}:=( {(r^2 + a^2)}/{\Delta}) dr$, and
$S_{\omega lm}(\theta, \phi):=
\Theta_{\wlm} (\theta) e^{i m \phi}$
is the spheroidal harmonics normalized as
\begin{equation}
\int 
d \theta d \phi \sin \theta
S_{\wlm}(\theta,\phi)^{\ast} 
S_{\omega l' m'}(\theta,\phi) = \delta_{l l'} \delta_{mm'}. 
\end{equation}
The superscript $\flat$ represents one of the four distinct boundary
conditions: ``up'', ``down'', ``in'' and ``out''.
The corresponding radial functions 
are defined by \cite{Gal'tsov:1982zz,Drasco:2005is}
\begin{eqnarray}\label{eq:uindef}
u_{\wlm}^{\rm in}(r_{\ast}) &=& 
\alpha_{\wlm}
\left\{ 
\begin{array}{ll} 
\tau_{\wlm} |p_{m\omega}|^{-1/2} e^{- i p_{m \omega} r_{\ast}},
 & \mbox{ ($r_{\ast} \to -\infty$), }  \\
|\omega|^{-1/2} 
\left[ e^{-i \omega r_{\ast}} + \sigma_{\wlm} e^{i\omega r_{\ast}} \right],   
 \qquad       &
\mbox{ ($r_{\ast} \to \infty$), }
\end{array} \right.
\nonumber \label{eq:uupdef} \\
u_{\wlm}^{\rm up}(r_{\ast}) &=& 
\beta_{\wlm}
\left\{ 
\begin{array}{ll}  
|p_{m\omega}|^{-1/2}
  \displaystyle\frac{\omega p_{m \omega}}{|\omega p_{m \omega}|}
  \left[ \mu_{\wlm} e^{ i p_{m \omega} r_{\ast}} 
  + \nu_{\wlm} e^{-i p_{m \omega} r_{\ast}} \right],
  & \mbox{ ($r_{\ast} \to -\infty$), } \\
|\omega|^{-1/2} e^{i \omega r_{\ast}},   
& \mbox{ ($r_{\ast} \to \infty$), }
\end{array} \right.
\nonumber \\
u^{{\rm out}}_{\wlm} (r_{\ast})  &=& u^{{\rm in} ~\ast}_{\wlm} (r_{\ast}), 
\nonumber \\
u^{{\rm down}}_{\wlm} (r_{\ast}) &=& u^{{\rm up} ~\ast}_{\wlm} (r_{\ast}),
\end{eqnarray}
respectively. Here 
$p_{m \omega}:= \omega - m a / (2 M r_{+} )$, 
$r_+ := M + \sqrt{M^2 - a^2}$,
and the superscript $\ast$ denotes the complex conjugate.
The coefficients $\alpha_{\wlm}$ and $\beta_{\wlm}$ are 
normalization constants. $\tau_{\wlm}$ and $\sigma_{\wlm}$ are 
complex transmission and reflection coefficients, respectively.
Both the complex coefficients
$\mu_{\wlm}$ and $\nu_{\wlm}$ are also defined by Eqs.~\eqref{eq:uupdef}.
In this work, we choose the normalizations of the ``in'' and  ``up''
solutions as
\begin{equation} \label{eq:u-normalization}
\alpha_{\wlm}^*=\alpha_\mwlm, \quad
\beta_{\wlm}^*=\beta_\mwlm,
\end{equation}
so that the invariance of the radial equation under the
transformation $(\omega, m)\rightarrow (-\omega, -m)$
implies the identity
\begin{equation}
u_{-\omega l -m}^\flat (r_{\ast})
= u_{\omega lm}^{\flat ~\ast} (r_{\ast}). \label{eq:id-u}
\end{equation}
From this identity (\ref{eq:id-u}), we find
\begin{equation} \label{eq:u-coeffs}
\tau_{\wlm}^*=\tau_\mwlm, \quad
\sigma_{\wlm}^*=\sigma_\mwlm, \quad
\mu_{\wlm}^*=\mu_\mwlm, \quad
\nu_{\wlm}^*=\nu_\mwlm.
\end{equation}
It should also be noted that we choose the phase of the normalization
of the angular function $\Theta_{\omega lm}(\theta)$ as 
\begin{equation}
\Theta_{-\omega l -m}^*(\theta)
= (-1)^m \Theta_{\omega lm}(\theta). \label{eq:id-Theta}
\end{equation}

The retarded Green function for the scalar field equation~(\ref{scalar-EOM})
is expressed in the factorized form of the mode functions 
\cite{Drasco:2005is}:
\begin{eqnarray}
G^{\rm (ret)}(x, x') &=& \frac{1}{16 \pi i} 
 \int_{-\infty}^{+\infty} d \omega 
\sum_{l = 0}^{+\infty} \sum_{m = -l }^{l}
\frac{\omega}{|\omega|}
\frac{1}{\alpha_{\wlm} \beta_{\wlm}} 
\left[ 
\pi^{\rm up}_{\wlm}(x) \pi^{\rm out\, *}_{\wlm}(x') H(r - r') 
+ 
\pi^{\rm in}_{\wlm}(x) \pi^{\rm down\, *}_{\wlm}(x') H(r' - r) 
\right]
\nonumber \\ &=&
\frac{1}{16 \pi i} \int_{-\infty}^{+\infty}
d \omega \sum_{l = 0}^{+\infty} \sum_{m = -l}^{l}
\frac{\omega}{|\omega|}
{1\over \alpha_\wlm\beta_\wlm}
e^{-i\omega(t-t')+im(\phi-\phi')}
\Theta_\wlm(\theta)\Theta^*_\wlm(\theta')
\cr
&&\hspace{2cm}
\times\Bigl[ 
R^{\rm up}_{\wlm}(r) R^{{\rm out}\,*}_{\wlm}(r') H(r-r')
+R^{{\rm out}\,*}_{\wlm}(r) R^{\rm up}_{\wlm}(r') H(r'-r)
\Bigr]
\label{G-retarded}
\end{eqnarray}
with the Heaviside function 
$H(x) := \int_{-\infty}^{x} \delta (y) dy$, and
$R_{\omega lm}^\flat(r) := 2 u_{\omega lm}^\flat/\sqrt{r^2+a^2}$.
In showing the second equality of Eq.~(\ref{G-retarded}),
we used the relations between the radial functions
$u_{\omega lm}^\flat(r_{\ast})$ in Eq.~(\ref{eq:uindef}).
This Green function leads the retarded solution as
\begin{equation}\label{s-retarded}
\Phi^{\rm (ret)}(x) =
\int d^4 x' \sqrt{-g(x')}\, G^{\rm (ret)}(x,x')\, {\cal T}(x')
= -q \int d \tau G^{\rm (ret)}[x, z(\tau)].
\end{equation}

For the later analysis, 
it is helpful to decompose the retarded field into 
a linear combination of the radiative (anti-symmetric) and
symmetric fields to be 
\begin{equation}\label{scalar-rad+sym}
\Phi^{\rm (ret)} (x) := \Phi^{\rm (rad)} (x) + \Phi^{\rm (sym)} (x), 
\end{equation}
where
\begin{equation}
\Phi^{\rm (rad)}(x) =
-q \int d \tau G^{\rm (rad)}[x, z(\tau)], \quad
\Phi^{\rm (sym)}(x) =
-q \int d \tau G^{\rm (sym)}[x, z(\tau)]
\end{equation}
with
\begin{eqnarray}
G^{\rm (rad)}(x,x') &=&
\frac{1}{2} \left[ G^{\rm (ret)}(x,x') - G^{\rm (adv)}(x,x') \right],
\label{eq:G-rad} \\
G^{\rm (sym)}(x,x') &=&
\frac{1}{2} \left[ G^{\rm (ret)}(x,x') + G^{\rm (adv)}(x,x') \right].
\label{eq:G-sym}
\end{eqnarray}
Here the advanced Green function, $G^{\rm (adv)}(x,x')$, is
derived from the retarded Green function as
\begin{equation}
G^{\rm (adv)}(x,x') = G^{\rm (ret)}(x',x).
\label{eq:G-adv}
\end{equation}
The radiative Green function can be rewritten as
\cite{Drasco:2005is}
\begin{eqnarray} \label{G-rad}
G^{\rm (rad)}(x,~x') &=& \frac{1}{32 \pi i} 
 \int_{-\infty}^{+\infty}
d \omega \sum_{l = 0}^{+\infty} \sum_{m = -l }^{l}
\frac{\omega}{|\omega|}
\left[ 
\frac{1}{|\alpha_{\wlm}|^2}
\pi^{\rm out}_{\wlm}(x) \pi^{\rm out\, *}_{\wlm}(x') 
+ 
\frac{\omega p_{m\omega}}{|\omega p_{m\omega}|}
\frac{|\tau_{\wlm}|^2}{|\beta_{\wlm}|^2}
\pi^{\rm down}_{\wlm}(x) \pi^{\rm down\, *}_{\wlm}(x') 
\right], \cr &&
\end{eqnarray}
and the Heaviside functions are eliminated.
On the other hand, we rewrite the symmetric Green function in the
following way.
Writing $G^{\rm (adv)}(x,x')$ in terms of the mode functions in a
manner similar to Eq.~(\ref{G-retarded}), replacing $(\omega, m)$
with $(-\omega, -m)$, and rearranging it with
Eqs.~(\ref{eq:u-normalization}), (\ref{eq:id-u}),
(\ref{eq:u-coeffs}), and (\ref{eq:id-Theta}), we find
\begin{eqnarray}
G^{\rm (adv)}(x,x') &=&
\frac{-1}{16 \pi i} 
 \int_{-\infty}^{+\infty}
d \omega \sum_{l = 0}^{+\infty} \sum_{m = -l }^{l}
\frac{\omega}{|\omega|}
{1\over \alpha^*_\wlm\beta^*_\wlm}
e^{-i\omega(t-t')+im(\phi-\phi')}
\Theta_\wlm(\theta)\Theta^*_\wlm(\theta')
\cr
&&\hspace{2cm}
\times\Bigl[ 
R^{\rm up\,*}_{\wlm}(r) R^{{\rm out}}_{\wlm}(r') H(r-r')
+R^{{\rm out}}_{\wlm}(r) R^{\rm up\,*}_{\wlm}(r') H(r'-r)
\Bigr]. \label{eq:G-adv2}
\end{eqnarray}
Substituting Eqs.~(\ref{G-retarded}) and
(\ref{eq:G-adv2}) into Eq.~(\ref{eq:G-sym}), we find that
the symmetric Green function is rewritten as
\begin{equation} \label{G-sym}
G^{\rm (sym)}(x,~x') 
= \frac{1}{16 \pi} \int_{-\infty}^{+\infty}
d \omega \sum_{l = 0}^{+\infty} \sum_{m = -l }^{l}
g_{\omega lm}^{\rm (sym)}(x, x'),
\end{equation}
where
\begin{eqnarray} \label{eq:gwlm-sym}
g_{\omega lm}^{\rm (sym)}(x, x') 
&=&
e^{-i\omega(t-t')+im(\phi-\phi')}
\Theta_\wlm(\theta)\Theta^*_\wlm(\theta')
\cr
&&
\times\Im\left[
\frac{\omega}{|\omega|}
\frac{1}{\alpha_\wlm\beta_\wlm }\Bigl[ 
R^{\rm up}_{\wlm}(r) R^{{\rm out}\,*}_{\wlm}(r') H(r-r')
+R^{{\rm out}\,*}_{\wlm}(r) R^{\rm up}_{\wlm}(r') H(r'-r)
\Bigr] \right],
\end{eqnarray}
and $\Im[X]$ represents the imaginary part of $X$.

\section{Adiabatic evolution of the Carter constant at resonance}
\label{sec:dIdt}

We denote the three constants of motion as 
$I^{i} := \{E,L,Q\}$. The time averaged rates of change
of $I^i$ can be re-expressed as
\begin{equation}\label{lambda-ave}
\left \langle \frac{d I^i}{dt} \right \rangle_t = 
\frac{1}{\Upsilon_t} \left \langle \frac{d I^i}{d \lambda} 
\right \rangle_{\lambda},
\end{equation}
with the definition of the $\lambda$ average, 
\footnote{
In the adiabatic approximation, we emphasize that the particle's motion 
can be treated as a geodesic motion in the background Kerr spacetime. 
The deviation from the geodesic due to the radiation reaction will affect 
the self-force by just a tiny amount. The influence of this deviation on 
the phase error will be $(O(M/\mu))^0$, which is beyond the accuracy that 
we are interested in here. 
Thus, taking the limit $T \to \pm \infty$ in Eq.~\eqref{lambda-ave2}
makes sense.}
\begin{equation}\label{lambda-ave2}
\langle F(\lambda) \rangle_{\lambda} := 
\lim_{T \to +\infty} \frac{1}{2 T}\int_{-T}^{+T} d \lambda'
F(\lambda'). 
\end{equation}
One key property of the adiabatic approximation developed
in Ref.~\cite{Mino:2005an} for off-resonance orbits is that
$\left \langle {dI^i}/{d\lambda} \right \rangle_\lambda$
can be evaluated by using the radiative solution of the
field equation~(\ref{scalar-EOM}) as
\begin{equation} \label{lambda-av-nonresonant}
\left \langle \frac{dI^i}{d\lambda} \right \rangle_\lambda
=
\left \langle
\frac{dI^i}{du^\alpha}
\bigl[ \Sigma(x) f^\alpha[\Phi^{\rm (rad)}(x)]
\bigr]_{x=z(\lambda)}
\right \rangle_\lambda,
\end{equation}
where $f^\alpha$ is a differential operator defined by
~\cite{Quinn:2000wa,Rosenthal:2003qr,Galley:2005tj}
\begin{equation}
f^\alpha[\Phi(x)] :=
q (g^{\alpha\beta}+u^\alpha u^\beta)
\nabla_\beta \Phi(x).
\end{equation}
Since the radiative field has no divergent behavior at
the location of the particle, this relation gives
$\left \langle {dI^i}/{d\lambda} \right \rangle_\lambda$
without any regularization procedure.
The final expressions for
$\left \langle {dI^i}/{d\lambda} \right \rangle_\lambda$
are given in terms of the asymptotic
amplitudes of the scalar field at infinity and on the
horizon~\cite{Drasco:2005is}.

The derivation of Eq.(\ref{lambda-av-nonresonant}) relies
on the assumption that there exists a simultaneous turning point
of the $r$- and $\theta$ -oscillations of the particle's motion.
In the case of a resonance orbit, however, the frequencies of the
$r$- and $\theta$ -oscillations are in a rational ratio, and hence
one cannot find a simultaneous turning point in general.
This means that the orbit depends on the non-vanishing
offset phase, $\Delta\lambda$. Therefore, the expressions for
$\left \langle {d I^i}/{dt} \right \rangle_t$ should be derived
without using the relation~(\ref{lambda-av-nonresonant}).

Incidentally, as for $\left \langle {d E}/{dt} \right \rangle_t$
and $\left \langle {d L}/{dt} \right \rangle_t$,
the balance argument can be used to relate them
to the fluxes of the energy and angular momentum 
to infinity or into the black hole horizon.
Therefore, one can conclude that the expressions for
$\left \langle {d E}/{dt} \right \rangle_t$ 
and $\left \langle {d L}/{dt} \right \rangle_t$
in the resonance case are the same as those in the
non-resonance case \cite{Flanagan:2012kg}.
On the other hand, there is no known flux 
composed of the perturbation field 
corresponding to the Carter constant, and hence we cannot
use the balance argument to derive the averaged rate of
change,  $\left \langle {dQ}/{dt} \right \rangle_t$.
We need to reformulate the calculation of
$\left \langle {dQ}/{dt} \right \rangle_t$
without using the relation~(\ref{lambda-av-nonresonant}).
In this section, we derive a formula for computing
$\left \langle {d Q}/{dt} \right \rangle_t$ in the resonance
case, by starting from the basic formula of the
self-force acting on a particle.

\subsection{Formal reduction of $\langle {dQ}/{dt} \rangle_t$
for a resonance orbit}
\label{subsec:simplify}
In contrast to $\left \langle {d E}/{dt} \right \rangle_t$ 
and $\left \langle {d L}/{dt} \right \rangle_t$, 
we need to directly evaluate the time derivative of the 
Carter constant defined in Eq.~\eqref{Carter-const}.
This is given by the self-force acting on the particle as 
\begin{equation}\label{def-dQdl}
\frac{dQ}{d \lambda} = 
2 \Sigma K_{\alpha \beta} u^{\alpha} a^{\beta}
\end{equation}
with the aid of the Killing tensor equation 
$\nabla_{( \gamma} K_{\alpha \beta )} = 0$.
Here, $a^{\alpha}$ is the four acceleration due to the scalar self-force 
with the retarded scalar field given by \cite{Detweiler:2002mi}
\begin{equation}\label{scalar-SF}
a^{\alpha} [z(\tau)] := 
f^\alpha[\Phi^{(R)}(x)]_{x = z(\tau)},
\end{equation}
where $\Phi^{(R)}$ is the so-called $R$-part of the retarded
solution, which satisfies the homogeneous equation of
Eq.~(\ref{scalar-EOM}). $\Phi^{(R)}$ is schematically expressed as
\begin{eqnarray}\label{R-ret}
\Phi^{(R)} (x) &:=&
\Phi^{\rm (ret)} (x)- \Phi^{(S)}(x)
\nonumber \\ &=&
\Phi^{\rm (rad)} (x)
+ \left\{ \Phi^{\rm (sym)} (x)- \Phi^{(S)}(x) \right\},
\end{eqnarray}
where $\Phi^{(S)}(x)$ is the $S$-part of the retarded field
\cite{Detweiler:2002mi},
which satisfies the same equation as the retarded (or symmetric)
solution.
It should be noted that, the difference between $\Phi^{\rm (sym)}$
and $\Phi^{(S)}$ is regular at the location of the particle
while each of them is singular.

Now plugging Eq.~(\ref{scalar-SF}) into Eq.~(\ref{def-dQdl}),
the time derivative of the Carter constant is rewritten by  
the $R$-part of the retarded field $\Phi^{(R)}(x)$:
\begin{equation}\label{Q-temp}
\frac{d Q}{d \lambda} = 
{2 Q q} \frac{d \Phi^{(R)}}{d \lambda}
+ {2 q \Sigma}K^{\alpha \beta} u_{\alpha} \nabla_{\beta} 
\Phi^{(R)}.
\end{equation}
Since the first term is a total derivative up to the leading
order in $q$, it does not contribute after taking a long-time
average.
Hence, we omit the first term in Eq.~(\ref{Q-temp}).
Substituting the explicit expression for $K_{\alpha \beta}$ 
into Eq.~\eqref{Q-temp}, and dropping the first term, we obtain
\begin{eqnarray}\label{dQdt}
\left \langle \frac{dQ}{dt} \right \rangle_t 
&=& 
- \frac {2 q}{ \Upsilon_t } 
\left \langle
\left\{ {\hat {\cal O}}_x \left[
\Sigma(x) \Phi^{(R)} (x)
\right] \right\}_{x=z(\lambda)} 
\right\rangle_{\lambda},
\end{eqnarray}
where ${\hat {\cal O}}_x$ is a differential operator
defined by
\begin{equation}
\hat{{\cal O}}_x
 :=  
V_{tr} (r) \partial_t + {V_{\phi r} (r)} \partial_{\phi} 
+ \frac{d r(\lambda)}{d \lambda} \partial_r.
\end{equation}
Following the decomposition of the retarded field in 
Eq.~\eqref{R-ret}, 
we also decompose $\left \langle {d Q}/{dt} \right \rangle_t$
into two parts as
\begin{eqnarray}
\left \langle \frac{dQ}{dt} \right \rangle_t 
&=&
\left \langle \frac{dQ}{dt} \right \rangle_t ^{\rm (rad)}
+ \left \langle \frac{dQ}{dt} \right \rangle_t ^{{\rm (sym}-S)}
\end{eqnarray}
with
\begin{eqnarray}
\left \langle \frac{dQ}{dt} \right \rangle_t^{\rm (rad)} 
&=&
\frac{2q^2}{\Upsilon_t}
\lim_{T\rightarrow\infty} \frac{1}{2T}
\int_{-T}^T d\lambda \int_{-\infty}^\infty d\lambda'
\left\{\hat{{\cal O}}_x \left[
\Sigma(x)\Sigma(x')G^{\rm (rad)}(x,x')
\right] \right\}_{x=z(\lambda), x'=z(\lambda')},
\label{eq:dQdt-rad} \\
\left \langle \frac{dQ}{dt} \right \rangle_t^{({\rm sym}-S)} 
&=&
\frac{2q^2}{\Upsilon_t}
\lim_{T\rightarrow\infty} \frac{1}{2T}
\int_{-T}^T d\lambda \int_{-\infty}^\infty d\lambda'
\left\{ \hat{{\cal O}}_x
\left[\Sigma(x)\Sigma(x')\left(
G^{\rm (sym)}(x,x') - G^{(S)}(x,x')
\right) \right] \right\}_{x=z(\lambda), x'=z(\lambda')},
\cr
&& \label{eq:dQdt-sym}
\end{eqnarray}
where $G^{(S)}(x,x')$ is the $S$-part of the retarded Green
function (see Eq.~(\ref{eq:G-Spart}) below).

We reduce Eqs.~(\ref{eq:dQdt-rad}) and (\ref{eq:dQdt-sym})
into more tractable expressions.
For this purpose, we consider a function of two variables,
$\tilde{G}(x,x')\equiv\Sigma(x)\Sigma(x')G(x,x')$.
Using the identity, 
\begin{equation}
\label{identity}
\frac{d}{d \lambda_r}
 \tilde{G}\left(\bar z(\lambda_r,\lambda_\theta), x'\right)
= 
\left[\left(
\frac{dr(\lambda_r)}{d\lambda_r}\partial_r 
+\frac{d\Delta t_r(\lambda_r)}{d\lambda_r} \partial_t 
+\frac{d\Delta \phi_r(\lambda_r)}{d\lambda_r} \partial_\phi
\right) \tilde{G}(x, x')
\right]_{x=\bar z(\lambda_r,\lambda_\theta)}, 
\end{equation}
we find
\begin{eqnarray}
\label{block1}
&& \hspace*{-0.7cm}
\lim_{T\to\infty}\frac{1}{2T}\int_{-T}^T d\lambda
\left\{ {\hat {\cal O}}_x \left[
 \tilde{G}(x, x')\right] \right\}_{x=z(\lambda)} \cr
&=&
\lim_{T\to \infty}{1\over 2T}\int_{-T}^T 
d\lambda_\theta 
\int_{-\infty}^{+\infty} d\lambda_r 
\, \delta(\lambda_r-\lambda_\theta+\Dl) 
\left\{ {\hat {\cal O}}_x \left[ \tilde{G}(x, x')
\right] \right\}_{x=\bar z(\lambda_r,\lambda_\theta)}\cr
&=&
\lim_{T\to \infty}{1\over 2T}\int_{-T}^T d\lambda_\theta
\int_{-\infty}^{+\infty} d\lambda_r  
\, \delta(\lambda_r - \lambda_\theta + \Dl)
\left\{
\left[ 
\left(
 \langle V_{tr}\rangle \partial_t 
 + \langle V_{\phi r}\rangle \partial_{\phi} \right)
  \tilde{G}(x, x') 
  \right]_{x=\bar z(\lambda_r,\lambda_\theta)}
+  \frac{d}{d \lambda_r} \left[
\tilde{G}\left(\bar z(\lambda_r,\lambda_\theta), x'\right)
\right]
\right\}
\nonumber \\ &=&
\left\langle
\left[ \left(
 \langle V_{tr}\rangle \partial_t 
 + \langle V_{\phi r}\rangle \partial_{\phi} \right)\tilde{G}(x, x') 
  \right]_{x=z(\lambda)}
\right \rangle_\lambda
-  \frac{d}{d (\Delta\lambda)} \big\langle
\tilde{G}\left[z(\lambda), x'\right]
\big\rangle_\lambda.
\end{eqnarray}
Here we insert the $\delta$-function, 
$\delta(\lambda_r - \lambda_\theta + \Dl)$, on the second
line to deal with the $r$- and $\theta$ -oscillations separately,
keeping the information on the offset phase.
In the last equality, we perform integration by parts with respect to 
$\lambda_r$ and replace 
$d\delta(\lambda_r-\lambda_\theta+\Dl)/d\lambda_r$
with $d\delta(\lambda_r-\lambda_\theta+\Dl)/d(\Dl)$. 
The differentiation $d/d(\Dl)$ can be extracted to the 
outside of the integral because the integrand depends on $\Dl$ 
only through this $\delta$-function.

As for $\langle dQ/dt \rangle_t^{\rm (rad)}$,
we arrive at the following expression
with the aid of Eq.(\ref{block1}):
\begin{eqnarray} \label{dQdt-3-rad}
\left \langle \frac{dQ}{dt} \right \rangle_t^{\rm (rad)}
&=&
\frac{2q^2}{\Upsilon_t} \int_{-\infty}^\infty d\lambda'
\Sigma[z(\lambda')] \left\langle
\left[ \Sigma(x) \Big(
\langle V_{tr} \rangle \partial_t
+ \langle V_{\phi r} \rangle \partial_\phi
\Big) G^{\rm (rad)}(x,z(\lambda'))
\right]_{x=z(\lambda)}
\right\rangle_\lambda
\nonumber \\ &&
- \frac{2q^2}{\Upsilon_t}
\int_{-\infty}^\infty d\lambda'
\left[ \Sigma(x')
\frac{d}{d(\Delta\lambda)}
\left\langle
\Sigma[z(\lambda)] G^{\rm (rad)}(z(\lambda), x')
\right\rangle_\lambda
\right]_{x'=z(\lambda')}.
\end{eqnarray}
We call attention to the fact 
that the $\Delta\lambda$-differentiation in the second
term operates only through $z(\lambda)$, not through $z(\lambda')$.

For $\left \langle {d Q}/{dt} \right \rangle_t^{({\rm sym}-S)}$,
we obtain a similar formula to the radiative part as
\begin{eqnarray}
\left \langle \frac{dQ}{dt} \right \rangle_t^{({\rm sym}-S)}
&=&
\frac{2q^2}{\Upsilon_t} \int_{-\infty}^\infty d\lambda'
\Sigma[z(\lambda')] \left\langle
\left[ \Sigma(x) \Big(
\langle V_{tr} \rangle \partial_t
+ \langle V_{\phi r} \rangle \partial_\phi
\Big) G^{({\rm sym}-S)}(x,z(\lambda'))
\right]_{x=z(\lambda)}
\right\rangle_\lambda
\nonumber \\ &&
- \frac{2q^2}{\Upsilon_t}
\int_{-\infty}^\infty d\lambda'
\left[ \Sigma(x')
\frac{d}{d(\Delta\lambda)} \left\langle
\Sigma[z(\lambda)] 
G^{({\rm sym}-S)}(z(\lambda), x')
\right\rangle_\lambda
\right]_{x'=z(\lambda')}, \label{eq:dQdt-sym1}
\end{eqnarray}
where
\[
 G^{({\rm sym}-S)}(x,x') :=
 G^{\rm (sym)}(x,x') - G^{(S)}(x,x').
\]
Using the following properties of $G^{({\rm sym}-S)}(x,x')$ 
(see Appendix \ref{app:simplify-dQdt-sym} for details),
\begin{eqnarray} 
G^{({\rm sym}-S)}(x,x') &=& G^{({\rm sym}-S)}(x',x),
\label{eq:Gsym-proper0} \\
G^{({\rm sym}-S)}(z(\lambda), z(\lambda')) &=&
G^{({\rm sym}-S)}(z(\lambda+\Lambda), z(\lambda'+\Lambda)),
\label{eq:Gsym-proper1} \\
\frac{\partial}{\partial t} G^{({\rm sym}-S)}(x,x')
&=&
-\frac{\partial}{\partial t'} G^{({\rm sym}-S)}(x,x'),
\quad
\frac{\partial}{\partial \phi} G^{({\rm sym}-S)}(x,x')
=
-\frac{\partial}{\partial \phi'} G^{({\rm sym}-S)}(x,x'),
\label{eq:Gsym-proper2}
\end{eqnarray}
we find that the first term in Eq.~(\ref{eq:dQdt-sym1}) vanishes, 
and then obtain 
\begin{eqnarray}
\left \langle \frac{dQ}{dt} \right \rangle_t^{({\rm sym}-S)}
&=&
- \frac{q^2}{\Upsilon_t} \frac{d}{d(\Delta\lambda)} \left[
\int_{-\infty}^\infty d\lambda'
\left\langle
\Sigma[z(\lambda)] \Sigma[z(\lambda')]
G^{({\rm sym}-S)}(z(\lambda),z(\lambda'))
\right\rangle_\lambda
\right]. \label{eq:dQdt-sym2}
\end{eqnarray}
In Eq.~(\ref{eq:dQdt-sym2}), the differentiation with respect to
$\Delta\lambda$ operates on both $z(\lambda)$ and $z(\lambda')$,
while, in the second term of Eq.~(\ref{eq:dQdt-sym1}),
it operates only on $z(\lambda)$. Since the part in the square
brackets in Eq.~(\ref{eq:dQdt-sym2}) is symmetric under the
exchange of $z(\lambda)$ and $z(\lambda')$, however, the
contribution from the $\Delta\lambda$-differentiation through
$z(\lambda')$ is the same as that through $z(\lambda)$.
This is the reason why the factor 2 in 
the expression of Eq.~(\ref{eq:dQdt-sym1})
disappears in Eq.~(\ref{eq:dQdt-sym2}).
The detailed derivation is also presented in 
Appendix \ref{app:simplify-dQdt-sym}.
The result that the $t$- and $\phi$-derivative parts in
Eq.~(\ref{eq:dQdt-sym1}) have no contribution, can be expected
from the fact that $\langle dE/dt \rangle_t$ and
$\langle dL/dt \rangle_t$ have no contribution from the
symmetric part of the Green function, which is shown by the
balance argument of the corresponding conserved currents.
We emphasize that, from the above expression, 
$\left \langle {dQ}/{dt} \right \rangle_t ^{({\rm sym}-S)}$
will not vanish in general. This means that
the secular change of the Carter constant would also be induced 
by the contribution from the symmetric part for the resonance orbit.

It should be noted that we cannot divide the integration in
Eq.~(\ref{eq:dQdt-sym2}) into symmetric and $S$-parts
because both $G^{\rm (sym)}[z(\lambda),z(\lambda')]$ 
and $G^{(S)}[z(\lambda),z(\lambda')]$ diverge
at the coincidence limit, $z(\lambda) \to z(\lambda')$.
In order to treat them separately,
we here introduce the point splitting regularization 
by displacing the orbits $z(\lambda)$ and $z(\lambda')$ as
\begin{equation}\label{dumbbell}
z^\mu(\lambda) \to z^\mu_+(\lambda):= 
z^\mu(\lambda) + \frac{\epsilon}{2}  \xi^\mu,
\qquad
z^\mu(\lambda') \to z^\mu_-(\lambda'):= 
z^\mu(\lambda') - \frac{\epsilon}{2} \xi^\mu
\end{equation}
with a small parameter $\epsilon \ll 1$ and the Killing field
\begin{equation}\label{helical}
\xi^\mu(\zeta) := \cos\zeta\, \xi_{(t)}^{\mu}  + 
(\Omega_\phi\cos\zeta-\Omega \sin\zeta)\, \xi_{(\phi)}^{\mu}
\end{equation}
as the direction of displacement. $\zeta$ is a parameter 
that specifies the choice of the Killing field,
$\Omega_\phi=\Upsilon_\phi/\Upsilon_t$ and
$\Omega=\Upsilon/\Upsilon_t$.
Under this point splitting regularization, we rewrite
Eq.~(\ref{eq:dQdt-sym2}) as
\begin{eqnarray}\label{eq:dQdt-sym3}
\left \langle \frac{dQ}{dt} \right \rangle_t^{({\rm sym}-S)}
& = & 
-\lim_{\epsilon\rightarrow 0}
\frac{q^2}{\Upsilon_t} \frac{d}{d(\Dl)} \left[
\Psi^{\rm (sym)}(\Delta\lambda; \epsilon, \zeta)
- \Psi^{(S)}(\Delta\lambda; \epsilon, \zeta)
\right].
\end{eqnarray}
The potential functions,
$\Psi^{\rm (sym)}(\Delta\lambda; \epsilon, \zeta)$
and $\Psi^{(S)}(\Delta\lambda; \epsilon, \zeta)$, are defined by
\begin{equation} \label{eq:dQdt-potential}
\Psi^{(X)}(\Delta\lambda; \epsilon, \zeta) =
\int_{-\infty}^{+\infty} d\lambda'\,
\left\langle
\Sigma[z(\lambda)] \Sigma[z(\lambda')]
G^{(X)}[z_+(\lambda),z_-(\lambda')] 
\right\rangle_\lambda,
\end{equation}
where the superscript $X$ is to be replaced with ``sym'' or ``$S$''
(notice that $\Sigma[z_{\pm}(\lambda)]=\Sigma[z(\lambda)]$ 
since $\Sigma(x)$ is a function of $r$ and $\theta$ ).
It is worth mentioning that the difference 
of the potential functions, 
$\Psi^{({\rm sym}-S)}(\Delta\lambda; \epsilon, \zeta) :=
\Psi^{\rm (sym)}(\Delta\lambda; \epsilon, \zeta) 
- \Psi^{(S)}(\Delta\lambda; \epsilon, \zeta)$, is even in 
the sign flip of $\Dl$. 
To show this, we replace $\lambda$ and $\lambda'$ 
with $-\lambda$ and $-\lambda'$. 
Under this replacement, 
the offset phase $\Dl$ (and the Killing vector $\xi^\mu(\zeta)$) 
also flips the sign. 
Hence, we find 
$\lim_{\epsilon \to 0}\Psi^{({\rm sym}-S)}(\Delta \lambda;\epsilon,\zeta)
 =\lim_{\epsilon \to 0}\Psi^{({\rm sym}-S)}(-\Delta \lambda;\epsilon,\zeta)$. 
This symmetry guarantees $\langle{dQ/ dt}\rangle_t^{({\rm sym}-S)}=0$ 
for $\Delta\lambda=0$ in general.

Using the mode decomposition of the radiative and symmetric
Green functions given in Eqs.~\eqref{G-rad} and \eqref{G-sym},
the expressions in Eqs.~\eqref{dQdt-3-rad}
and \eqref{eq:dQdt-sym3} (with Eq.~(\ref{eq:dQdt-potential}))
can be recast into more practical expressions.
We will show them in the following subsections.

\subsection{Further reduction of 
$\left \langle {d Q}/{dt} \right \rangle_t^{\rm (rad)}$: 
the radiative part}
\label{subsec:dQdt-R}
%
%
Flanagan~{\etal} \cite{Flanagan:2012kg} have already written
down a little more practical expression
for the radiative part of the long-time averaged value of the rate of
change of the Carter constant,
$\left \langle {d Q}/{dt} \right \rangle_t^{\rm (rad)}$,
in the gravitational case. Here, we briefly rederive the same result 
in the scalar case in a slightly different manner.

Our first task is to rewrite Eq.~\eqref{dQdt-3-rad} 
in terms of the mode functions given in
Eq.~\eqref{complete-mode}.
With the aid of the radiative Green function \eqref{G-rad} 
expressed as a sum of mode functions, we have 
\begin{eqnarray}\label{dQdt-wlm}
\left \langle \frac{dQ}{dt} \right \rangle_t ^{\rm (rad)} 
& = &
\frac{q^2}{16 \pi i \Upsilon_t}
 \int_{-\infty}^{+\infty}
d \omega \sum_{l = 0}^{+\infty} \sum_{m = -l }^{l}
\frac{\omega}{|\omega|}
\cr
&&\times
\Biggl[ 
{1 \over |\alpha_{\wlm}|^2}
\left\{
\int_{-\infty}^{+\infty}d \lambda'
\tilde{\pi}^{{\rm out}\,*}_{\wlm}[z(\lambda')] 
\right\}
\left(-i \omega\langle V_{tr}\rangle + im \langle V_{\phi r} \rangle 
- \frac{d}{d(\Delta\lambda)} \right) 
\langle \tilde \pi^{\rm out}_{\wlm}[z(\lambda)] \rangle_\lambda
\cr && + 
\frac{\omega p_{m\omega}}{|\omega p_{m\omega}|}
\frac{|\tau_{\wlm}|^2}{|\beta_{\wlm}|^2}
\left\{ \int_{-\infty}^{+\infty} d \lambda'
\tilde\pi^{{\rm down}\,*}_{\wlm}[z(\lambda')] \right\}
\left(-i \omega\langle V_{tr}\rangle + im \langle V_{\phi r} \rangle 
- \frac{d}{d(\Delta\lambda)} \right) 
\langle \tilde\pi^{\rm down}_{\wlm} [z(\lambda)]  \rangle_\lambda
\Biggr], \cr&&
\end{eqnarray}
where, for brevity, we have introduced
$\tilde\pi^{\flat}_{\wlm}(x) := \Sigma(x)\pi^{\flat}_{\wlm}(x)$,
and also replaced the partial differentiations with respect to
$t$ and $\phi$ with $-i\omega$ and $im$, respectively.
We can write $\tilde\pi^{{\flat}}_{\wlm}[z(\lambda)]$ as
\begin{eqnarray}
\label{Z-wlm}
\tilde\pi^{{\flat}}_{\wlm}[z(\lambda)]
&=& \, {\cal J}_{\wlm}^{\flat\, *} (\lambda - \Delta \lambda, \lambda )
e^{-i \lambda (\omega \Upsilon_t  - m \Upsilon_{\phi})}
\end{eqnarray}
with a function of two time variables, $\lambda_r$ and 
$\lambda_\theta$: 
\begin{equation}\label{J-wlm}
{\cal J}_{\wlm}^{\flat} (\lambda_r, \lambda_{\theta}) 
:= 
 \Sigma(x)\pi^{\flat\,*}_{\wlm}(x)
 \Bigr|_{x=\Delta \bar z(\lambda_r,\lambda_\theta)},
\end{equation}
where $\Delta \bar z(\lambda_r,\lambda_\theta)$ is 
the oscillating part of the orbit defined in 
Eq.~\eqref{delta-barz}.
Since the function ${\cal J}_{\wlm}^{\flat} (\lambda_r, \lambda_\theta)$
is biperiodic with the frequencies $\Upsilon_r$ and 
$\Upsilon_\theta$ for $\lambda_r$ and $\lambda_\theta$, respectively, 
it can be expanded in the Fourier series as
\begin{eqnarray}\label{w-mkn}
{\cal J}^{\flat}_{\wlm} (\lambda_r, \lambda_\theta)
 &=&
\sum_{n_r=-\infty}^\infty
\sum_{n_\theta=-\infty}^\infty
\tilde{{\cal J}}^{\flat}_{l m n_r n_\theta}(\omega)
e^{- i (n_r\Upsilon_r\lambda_r 
+ n_\theta\Upsilon_\theta \lambda_\theta)},
\end{eqnarray}
where
\begin{eqnarray}
\tilde{{\cal J}}^{\flat}_{l m n_r n_\theta}(\omega)
&:= &
\frac{1}{\Lambda_r} \int_0^{\Lambda_r} 
d \lambda_r 
\frac{1}{\Lambda_\theta} \int_0^{\Lambda_\theta} 
d \lambda_\theta \, 
{\cal J}^{\flat}_{\wlm} (\lambda_r,\lambda_\theta)
e^{i 
(n_r\Upsilon_r\lambda_r + n_\theta\Upsilon_\theta\lambda_\theta)}.
\nonumber
\end{eqnarray}
By using Eq.~(\ref{Z-wlm}) with Eq.~(\ref{w-mkn}), we express
the $\lambda'$-integrals in Eq.~(\ref{dQdt-wlm}) as
\begin{eqnarray}
\int_{-\infty}^\infty d\lambda'
\tilde{\pi}_{\omega lm}^{\flat\, *}[z(\lambda')]
&=&
\sum_{n_r=-\infty}^\infty
\sum_{n_\theta=-\infty}^\infty
\frac{2\pi}{\Upsilon_t}
\tilde{{\cal J}}_{lmn_r n_\theta}^{\flat}(\omega)
e^{i n_r \Upsilon_r \Delta\lambda}
\delta(\omega-\omega_{mn_r n_\theta})
\nonumber \\ &=&
\sum_{N=-\infty}^\infty
Z_{lmN}^\flat(\Delta\lambda)
\delta(\omega-\omega_{mN}),
\label{eq:omega-discretized}
\end{eqnarray}
where we introduce 
\begin{eqnarray}
\omega_{m n_r n_\theta} &:=&
\Upsilon_t^{-1} \left(
m\Upsilon_\phi + n_r\Upsilon_r + n_\theta\Upsilon_\theta
\right)
=
m\Omega_\phi + ( j_r n_r + j_\theta n_\theta ) \Omega,
\label{discrete-w} \\
\omega_{mN} &:=&
m\Omega_\phi + N \Omega,
\label{discrete-wmN}
\end{eqnarray}
and
\begin{eqnarray}
Z_{lmN}^{\flat}(\Delta\lambda) &:=&
\frac{2\pi}{\Upsilon_t} \sum_{(n_r,n_\theta)\in\hat{N}}
\tilde{{\cal J}}_{lmn_r n_\theta}^{\flat}(\omega_{mn_r n_\theta})
e^{i n_r \Upsilon_r \Delta\lambda},
\end{eqnarray}
with
\begin{equation} \label{eq:hatN}
\hat{N} :=
\left\{
(n_r, n_\theta) |
j_r n_r + j_\theta n_\theta = N
\right\}.
\end{equation}
Here it should be noted that $N$ denotes an integer,
and that we rearranged the order of summation over $(n_r, n_\theta)$
in the second equality in Eq.~(\ref{eq:omega-discretized}).

Owing to the $\delta$-function in Eq.~(\ref{eq:omega-discretized}),
we can replace the Fourier integral in Eq.~(\ref{dQdt-wlm})
with the discrete Fourier series. Hence, we can also replace
$\omega$ in $\langle \tilde{\pi}_{\omega_{mN}lm}^{\flat}
[z(\lambda)] \rangle_\lambda$ with $\omega_{mN}$, 
and then
\begin{eqnarray}
\langle \tilde{\pi}_{\omega_{mN}lm}^{\flat}
[z(\lambda)] \rangle_\lambda
&=&
\lim_{T\rightarrow\infty}\frac{1}{2T}\int_{-T}^T d\lambda
\sum_{n_r, n_\theta}
\tilde{{\cal J}}_{lmn_r n_\theta}^{\flat\, *}(\omega_{mN})
e^{i\lambda
  \left( n_r \Upsilon_r + n_\theta \Upsilon_\theta
  - N \Upsilon \right)
  - in_r\Upsilon_r\Delta\lambda}
\nonumber \\ &=&
\sum_{n_r, n_\theta}
\tilde{{\cal J}}_{lmn_r n_\theta}^{\flat\, *}(\omega_{mN})
e^{-in_r\Upsilon_r\Delta\lambda}
\frac{1}{\Lambda}\int_{-\Lambda/2}^{\Lambda/2} d\lambda
e^{i\lambda\Upsilon
  \left( n_r j_r + n_\theta j_\theta - N \right) }
\nonumber \\ &=&
\frac{\Upsilon_t}{2\pi}
Z_{lmN}^{\flat\, *}(\Delta\lambda).
\label{eq:t-average-pi}
\end{eqnarray}
In the second equality, we used the periodicity of the
integrand with the period $\Lambda$.

Finally, using Eqs.~(\ref{eq:omega-discretized}) and
(\ref{eq:t-average-pi}), we can reduce Eq.~(\ref{dQdt-wlm}) to
\begin{eqnarray}\label{keyresults}
\left \langle \frac{dQ}{dt} \right \rangle_t ^{\rm (rad)}
\!\!\!\! 
& = \!\!& 
-\frac{q^2}{32 \pi^2 } 
\sum_{\sharp} \frac{\omega_{mN}}{|\omega_{mN}|}
\Biggl[
\left(\omega_{mN}\langle V_{tr}\rangle 
 - m \langle V_{\phi r} \rangle \right)
\left\{ 
\frac{|Z^{\rm out}_{lmN}(\Delta\lambda)|^2}
{\left| \alpha_{\omega_{mN}lm}\right|^2} 
+
\frac{\omega_{mN} p_{m\omega_{mN}}}
{\left| \omega_{mN} p_{m\omega_{mN}} \right|}
\frac{\left| \tau_{\omega_{mN}lm} \right|^2}
{\left| \beta_{\omega_{mN}lm} \right|^2}
|Z^{\rm down}_{lmN}(\Delta\lambda)|^2 
\right\}
\cr &&
- \Upsilon_r \left\{
\frac{1}{\left| \alpha_{\omega_{mN}lm}\right|^2}
Z_{lmN}^{\rm out}(\Delta\lambda)
{\cal Y}_{lmN}^{{\rm out} *}(\Delta\lambda)
+
\frac{\omega_{mN} p_{m\omega_{mN}}}
{\left| \omega_{mN} p_{m\omega_{mN}} \right|}
\frac{\left| \tau_{\omega_{mN}lm} \right|^2}
{\left| \beta_{\omega_{mN}lm} \right|^2}
Z_{lmN}^{\rm down}(\Delta\lambda)
{\cal Y}_{lmN}^{{\rm down} *}(\Delta\lambda)
\right\}
\Biggr]
\nonumber \\ & = &
2 \left(\left \langle \frac{dE}{dt} \right \rangle_t
  \langle V_{tr}\rangle 
 - \left \langle \frac{dL}{dt} \right \rangle_t
    \langle V_{\phi r} \rangle \right)
\cr &&
+ \frac{ q^2 \Upsilon_r }{32 \pi^2 } \sum_{\sharp}
\frac{\omega_{mN}}{|\omega_{mN}|}
\left\{
\frac{1}{\left| \alpha_{\omega_{mN}lm}\right|^2}
Z_{lmN}^{\rm out}(\Delta\lambda)
{\cal Y}_{lmN}^{{\rm out} *}(\Delta\lambda)
+
\frac{\omega_{mN} p_{m\omega_{mN}}}
{\left| \omega_{mN} p_{m\omega_{mN}} \right|}
\frac{\left| \tau_{\omega_{mN}lm} \right|^2}
{\left| \beta_{\omega_{mN}lm} \right|^2}
Z_{lmN}^{\rm down}(\Delta\lambda)
{\cal Y}_{lmN}^{{\rm down} *}(\Delta\lambda)
\right\} \cr &&
\end{eqnarray}
with $\sum_{\sharp} 
:= \sum_{l = 0}^{+\infty} \sum_{m = -l }^{l} 
\sum_{N = -\infty}^{+\infty}$ and
\begin{equation}
{\cal Y}_{lmN}^\flat(\Delta\lambda)
:=
\frac{1}{i\Upsilon_r}\frac{d}{d(\Delta\lambda)}
Z_{lmN}^\flat(\Delta\lambda)
=
\frac{2\pi}{\Upsilon_t}
\sum_{(n_r, n_\theta)\in\hat{N}}
n_r \tilde{\cal J}_{lmn_r n_\theta}^\flat
e^{i n_r \Upsilon_r \Delta\lambda}.
\end{equation}
In the second equality, we used the expressions
of $\left \langle {d E}/{dt} \right \rangle_t$ 
and $\left \langle {d L}/{dt} \right \rangle_t$
\cite{Drasco:2005is},
\begin{eqnarray}
\left \langle \frac{d}{dt} 
\left\{ \begin{array}{l} 
E \\
L   \\
\end{array} \right\}
\right \rangle_t 
= 
- \frac{q^2}{64 \pi^2 }
\sum_\sharp
\left\{ \begin{array}{l} 
\omega_{mN} \\
m   \\
\end{array} \right\}
\frac{\omega_{mN}}{|\omega_{mN}|}
\left[
\frac{|Z^{\rm out}_{lmN}(\Delta\lambda)|^2}
{\left| \alpha_{\omega_{mN}lm}\right|^2} 
+
\frac{\omega_{mN} p_{m\omega_{mN}}}
{\left| \omega_{mN} p_{m\omega_{mN}} \right|}
\frac{\left| \tau_{\omega_{mN}lm} \right|^2}
{\left| \beta_{\omega_{mN}lm} \right|^2}
\left|Z^{\rm down}_{lmN}(\Delta\lambda)\right|^2 
\right].
\label{eq:dcalEdt2}
\end{eqnarray}
The contribution from the radiative field 
in Eq.~\eqref{keyresults}
is nothing but the scalar analogue of
$\left \langle {dQ}/{dt} \right \rangle^{\rm (rad)}$ 
obtained by Flanagan~{\etal}~\cite{Flanagan:2012kg}.
We note that the above expression for the radiative part is also valid 
for the off-resonance case if we interpret the summation over 
$N$ as meaning the summation over independent frequencies in this case.

\subsection{Practical formula of 
$\left \langle {d Q}/{dt} \right \rangle_t^{({\rm sym}-S)}$:
the regularized symmetric part}
\label{subsec:dQdt-Sym}
%
To obtain a more practical expression for
$\left \langle {d Q}/{dt} \right \rangle_t^{({\rm sym}-S)}$, 
we rewrite the potential functions of the symmetric and
$S$-parts given in Eq.~(\ref{eq:dQdt-potential}), respectively.
The symmetric part can be treated in a similar manner to the
radiative part discussed in Sec.~\ref{subsec:dQdt-R}, though we cannot
completely separate the averages over $\lambda$ and $\lambda'$
at the level of each mode.
Substituting Eq.~(\ref{G-sym}) (with Eq.~(\ref{eq:gwlm-sym})) into
the potential function of the symmetric part in Eq.~
(\ref{eq:dQdt-potential}), we obtain
\begin{eqnarray} \label{eq:Psi-sym}
\Psi^{\rm (sym)}(\Delta\lambda; \epsilon, \zeta)
&=&
\frac{1}{16\pi}\int_{-\infty}^\infty \!\!\!\! d\omega \sum_{lm}
e^{-i\omega\epsilon\cos\zeta
+ im\epsilon(\Omega_\phi\cos\zeta-\Omega\sin\zeta)}
\cr && \times
\int_{-\infty}^\infty \!\!\!\! d\lambda' \left\langle
e^{-i(\lambda-\lambda')(\omega\Upsilon_t-m\Upsilon_\phi)}
\tilde{g}_{\omega lm}^{\rm (sym)}
[\Delta\bar{z}(\lambda, \lambda-\Delta\lambda),
\Delta\bar{z}(\lambda', \lambda'-\Delta\lambda)]
\right\rangle_\lambda,
\end{eqnarray}
where
\[
\tilde{g}_{\omega lm}^{\rm (sym)}(x, x')
:= \Sigma(x) \Sigma(x') g_{\omega lm}^{\rm (sym)}(x, x').
\]
Since $\tilde{g}_{\omega lm}^{\rm (sym)}
[\Delta\bar{z}(\lambda, \lambda-\Delta\lambda),
\Delta\bar{z}(\lambda', \lambda'-\Delta\lambda)]$ is a
periodic function with period $\Lambda$,
we find that the $\lambda'$-integral in Eq.~(\ref{eq:Psi-sym})
produces $\delta(\omega-\omega_{mN})$,
and therefore we can discretize the Fourier integral as
\begin{eqnarray} \label{eq:Psi-sym-discretized}
\Psi^{\rm (sym)}(\Delta\lambda; \epsilon, \zeta)
&=&
\frac{1}{8\Upsilon_t} \sum_\sharp
e^{-i\Omega (\epsilon_1 N + \epsilon_2 m)}
\frac{1}{\Lambda} \int_0^\Lambda \!\!\!\! d\lambda'
\left\langle
e^{-iN\Upsilon(\lambda-\lambda')}
\tilde{g}_{\omega_{mN} lm}^{\rm (sym)}
[\Delta\bar{z}(\lambda, \lambda-\Delta\lambda),
\Delta\bar{z}(\lambda', \lambda'-\Delta\lambda)]
\right\rangle_\lambda,
\end{eqnarray}
where $\epsilon_1=\epsilon\cos\zeta$, and
$\epsilon_2=\epsilon\sin\zeta$.
In the above equation, the long-time average over $\lambda$ can be
replaced by an average over one period because the part in the angle
brackets is periodic with period $\Lambda$. 
For later convenience, in subtracting the 
the potential function of the $S$-part,
$\Psi^{(S)}(\Delta\lambda; \epsilon, \zeta)$, from
$\Psi^{\rm (sym)}(\Delta\lambda; \epsilon, \zeta)$, we rewrite
Eq.~(\ref{eq:Psi-sym-discretized}) by rearranging the order of
the summation with respect to $l$ and $m$ as
\begin{equation} \label{eq:Psi-sym-discretized2}
\Psi^{\rm (sym)}(\Delta\lambda; \epsilon, \zeta)
=
\sum_{N=-\infty}^\infty \sum_{m=-\infty}^\infty
e^{-i\Omega (\epsilon_1 N + \epsilon_2 m)}
\Psi_{mN}^{\rm (sym)}(\Delta\lambda).
\end{equation}
Here we define the mode decomposition of 
$\Psi^{\rm (sym)}(\Delta\lambda; \epsilon, \zeta)$ as 
\begin{equation} \label{eq:Psi-sym_mode}
\Psi_{mN}^{\rm (sym)}(\Delta\lambda) :=
\frac{1}{8\Upsilon_t} \sum_{l=|m|}^\infty
\frac{1}{\Lambda^2} \int_0^\Lambda \!\!\!\! d\lambda
\int_0^\Lambda \!\!\!\! d\lambda'
e^{-iN\Upsilon(\lambda-\lambda')}
\tilde{g}_{\omega_{mN} lm}^{\rm (sym)}
[\Delta\bar{z}(\lambda, \lambda-\Delta\lambda),
\Delta\bar{z}(\lambda', \lambda'-\Delta\lambda)].
\end{equation}
Equations~(\ref{eq:Psi-sym-discretized2}) and
(\ref{eq:Psi-sym_mode}) can be interpreted as 
the Fourier expansion in 
the $(\epsilon_1, \epsilon_2)$-space
and the Fourier coefficients, respectively. The point is
that each $(m, N)$ Fourier mode is finite, which can be
calculated numerically.

Next, we derive $\Psi^{(S)}(\Delta\lambda; \epsilon, \zeta)$.
Using half the squared geodesic distance between $x$ and $x'$,
$\sigma(x,x')$, the $S$-part Green function defined in the local
convex neighborhood of the particle's position is written as 
\begin{equation} \label{eq:G-Spart}
G^{(S)}(x,x') = 
-\frac{1}{8\pi}
\left[U(x,x')\delta(\sigma(x,x')) {-} V(x,x') \theta(\sigma(x,x'))\right]
\end{equation}
with
\begin{equation}
U(x,x') = 1 + O(|x-x'|^3/L^3),
\end{equation}
where $L$ is the scale of curvature of the background spacetime, 
and $V(x,x')$ is a regular function in the coincidence limit 
$x \to x'$ (see, \eg, Ref.~\cite {Poisson:2011nh} for
more concrete definitions of these quantities). 
The integrand containing $V(x,x')$ contributes as the 
finite value to $\Psi^{(S)}(\Dl; \epsilon, \zeta)$ since
the $\lambda'$ integral has its support only in a short interval
around $\lambda' \approx \lambda$. 
In the coincidence limit, this interval shrinks to zero, 
and thus this contribution also vanishes. 
Then, we need to consider only the term including $U(x, x')$.
Abbreviating the argument $\Dl$,
the argument of the $\delta$-function 
is expanded as
\begin{eqnarray}
\label{expansion2}
\sigma[z_+(\lambda),z_-(\lambda')] 
&=& \frac{1}{2} g_{\mu\nu} [z(\bar\lambda)]
\left\{ 
z^{\mu}(\lambda) - z^{\mu}(\lambda')  +  \epsilon \xi^{\mu} 
\right\}
\left\{
z^{\nu} (\lambda) - z^{\nu}(\lambda') +  \epsilon \xi^{\nu} 
\right\} + O(\epsilon^4)~\cr
&=& {\frac{1}{2}} g_{\mu\nu}[ z(\bar\lambda)] 
\left\{ 
\Sigma[z(\bar\lambda)] u^{\mu} (\bar \lambda) 
          \delta \lambda + \epsilon \xi^{\mu} \right\} 
\left\{
\Sigma[z(\bar\lambda)] u^{\nu} (\bar \lambda) 
          \delta \lambda + \epsilon \xi^{\nu} \right\}  
+ O(\epsilon^4),
\end{eqnarray}
where we have introduced the new variables 
\begin{equation}
{
\delta \lambda := (\lambda - \lambda'),
}
\qquad
\bar \lambda := {1\over 2}(\lambda + \lambda'),
\end{equation}
and assumed $\delta\lambda = O(\epsilon)$ since the solution of 
$\sigma=0$ for $\delta\lambda$ has the same behavior.
We stress that the $O(\epsilon^3)$ term is absent since 
$\sigma$ is unchanged under the transformation 
$\epsilon \to -\epsilon$ and $\delta\lambda\to -\delta\lambda$.
With the aid of the above expansion for $\sigma$, 
we expand both $ \Sigma[z(\lambda)]$
and $\Sigma[z(\lambda')]$ in terms of $\epsilon$ as well,
and eventually obtain 
\begin{equation}
\label{singular-U}
\Psi^{(S)}(\Delta \lambda; \epsilon, \zeta) = 
{-}
\frac{\psi(\zeta)}{\epsilon} + O(\epsilon)
\end{equation}
with 
\begin{equation}
\psi(\zeta):= \frac{1}{4 \pi \Lambda}
\int_{0}^\Lambda d\bar \lambda
\frac{\Sigma(x)} 
{\sqrt{(g_{\mu\nu}+u_{\mu}u_{\nu}) \xi^\mu \xi^\nu
}}
\Biggr\vert_{x=z(\bar \lambda)},
\end{equation}
where we replace the long-time average over
$\bar\lambda$ by the average over one period because the
integrand is periodic with period $\Lambda$.
We note that $\psi(\zeta)$ depends on $\zeta$ 
through $\xi^{\mu}$.
This is the only contribution of the $S$-part 
that remains in the limit $\epsilon \to 0$.
Since $(g_{\mu\nu}+u_{\mu}u_{\nu})$ is a
spacelike projection operator, the expression inside the square root
is positive semi-definite. 

To subtract the $S$-part of the potential function 
from the symmetric part mode by mode, we calculate the $(m, N)$-modes of
$\Psi^{(S)}(\Delta\lambda; \epsilon, \zeta)$ as
\begin{eqnarray}
\Psi^{(S)}_{mN} (\Dl)
&=&
{\Omega^2\over 4\pi^2}
\int_{- {\pi \over \Omega} }^{\pi\over \Omega} 
d\epsilon_1 \int_{- {\pi \over \Omega} }^{\pi\over \Omega} 
d\epsilon_2 e^{i \Omega (\epsilon_1 N+\epsilon_2 m)} 
\Psi^{(S)}(\Dl; \epsilon, \zeta)
\nonumber \\ &=&
{ - \frac{\Omega^2}{4\pi^2}}
\int_{-{\pi\over \Omega}}^{\pi\over \Omega} d\epsilon_1 
\int_{-{\pi\over \Omega}}^{\pi\over \Omega} d\epsilon_2 
e^{i \Omega (\epsilon_1 N+\epsilon_2 m)} 
\left( \psi(\zeta)\over \sqrt{\epsilon_1^2 + \epsilon_2^2} \right)
\nonumber \\ &=&
{ - \frac{\Omega^2}{4\pi^2}}
\left[
\int_{-{\pi\over 4}}^{\pi\over 4}
{d\zeta \over \cos\zeta} \int_{-{\pi\over \Omega}}^{\pi\over \Omega} 
d\epsilon_1 e^{i\epsilon_1 \Omega (N+m\tan\zeta)}\psi(\zeta)
-\int_{\pi\over 4}^{3\pi\over 4}
{d\zeta\over\sin\zeta} \int_{-{\pi\over \Omega}}^{\pi\over \Omega} 
d\epsilon_2\, e^{i\epsilon_2 \Omega (N\cot\zeta+m)}\psi(\zeta)\right]
\nonumber \\ &=&
{\Omega\over 2\pi^2}
\left[
(-1)^m \int_{{\pi\over 4}}^{3\pi \over 4}
{d\zeta} 
\frac{\sin(\pi N\cot\zeta) \psi(\zeta)}{N\cos\zeta + m\sin\zeta}
- (-1)^N \int_{-{\pi\over 4}}^{\pi \over 4}
{d\zeta} 
\frac{\sin(\pi m\tan\zeta) \psi(\zeta)}{N \cos\zeta + m \sin \zeta}
\right]. \label{singular-Psi}
\end{eqnarray}
This expression is a double integral with respect to 
$\bar\lambda$ and $\zeta$, and is finite. 
Since $\psi(\zeta)$ 
is independent of $m$ and $N$, once we obtain
$\psi(\zeta)$ as a function of $\zeta$ by performing 
the $\bar\lambda$-integral, the remaining integral over $\zeta$ 
for each pair of $m$ and $N$ is just a single integral.

We stress that subtracting Eq.~(\ref{singular-Psi})
from Eq.~(\ref{eq:Psi-sym_mode})
is a variant of the so-called mode sum regularization 
\cite{Barack:2009ux, Barack:2001gx} 
developed in the context of self-force, but our proposal is suitable
for the decomposition in terms of the {\it spheroidal harmonics}.
We do not have to re-expand the perturbation using the spherical
harmonics, which is usually utilized.
After subtracting this $S$-part contribution from 
$\Psi_{mN}^{\rm (sym)} (\Dl)$ mode by mode, 
one should be able to take the summations over $m$ and $N$ 
without any divergence. 
The convergence of these summations will be even accelerated 
by fitting the asymptotic form as usual. 
We will report the result of the explicit convergence test of 
$\Psi_{mN}^{\rm (sym)} (\Dl)-\Psi_{mN}^{(S)} (\Dl)$
in our next publication.

\section{Summary and Future directions}
\label{sec:Conclusion}
In this paper, we have derived the long-time averaged rate of change 
of the Carter constant, $\langle dQ/dt \rangle_t$,
for a self-interacting scalar charge within the adiabatic regime,
which is also applicable to a resonant inspiral.
The essential point is that, by contrast to off-resonance orbits, 
a resonance orbit depends on the offset phase, $\Dl$,
which is the difference between the times when the $r$- 
and $\theta$ -oscillations reach their minimum values.
While this offset phase is identically zero for an off-resonance
orbit, it takes a non-zero value for a resonance orbit, 
and also evolves slowly compared to the time scale of 
the orbital motion.
Therefore, we need to concern ourselves with the evolution of not only
the three constants of motion but also $\Delta \lambda$. 
The existence of a non-zero $\Delta\lambda$ requires a direct
computation of the self-force acting on the charged particle
even for calculating the averaged rate of change of the Carter
constant. This means that $\langle dQ/dt \rangle_t$ for a resonance
orbit is no longer determined solely by the radiative field 
$\Phi^{\rm (rad)}(x)$ alone, and that  we need to consider the
contribution from the symmetric field $\Phi^{\rm (sym)}(x)$ 
as well.

As for the radiative part $\langle dQ/dt \rangle_t^{\rm (rad)}$,
we have derived a practical expression in a similar manner to the 
off-resonance cases with slight modification, and found that
our result is equivalent to the scalar analogue of the long-time
averaged rate for the gravitational case shown
in Ref.~\cite{Flanagan:2012kg}.
On the other hand, the symmetric part is more complicated
than the radiative one because of the singular behavior
of the symmetric field at the location of the particle.
We have proposed a method of subtracting the singular behavior, 
the so-called $S$-part, from the symmetric part via the point splitting
regularization in the Killing directions.
The key point is that it
is compatible with the Teukolsky formalism, which makes the
calculation much simpler than the direct self-force calculation.
While there is an expectation that 
$\langle {dQ}/{dt} \rangle_t^{({\rm sym}-S)}$ might 
eventually become suppressed, which is suggested by post-Newtonian 
analysis~\cite{Flanagan:2012pc},
this conjecture should be tested in a fully general relativistic 
analysis.
Since the numerical code that calculates the long-time average of the 
constants of motion for generic off-resonance orbits around 
the Kerr black hole is well developed \cite{Drasco:2005kz, Fujita:2009us}, 
there is in principle no obstacle to numerically testing the conjecture.
The code is now under development and 
the result will be reported in our next publication.

To avoid unnecessary confusion over various works,
we comment on the following two points related to our current work.
The first point is the difference between the long-time average 
and the  phase space average for a resonance orbit, 
discussed in Refs.~\cite{Hinderer:2008dm,Flanagan:2010cd}.
Unlike the long-time average,
the phase space average for a resonance orbit means  
that the averaging is further performed with respect 
to the slowly evolving variable $\Dl$, which is an important 
control variable of the resonance orbit.
The phase space average eliminates all dependence of $\Dl$ 
both in $\langle dQ/dt \rangle_t^{({\rm rad})}$ as well as 
$\langle dQ/dt \rangle_t^{({\rm sym}-S)}$, 
and thus it is likely to miss the important contribution to the 
accumulated orbital phase correction that would exceed $O(1)$ radians 
(see the discussion in Appendix~\ref{app:evolution}).
For this reason, we think that only the long-time averaged rate 
of change of constants of motion can be 
a good candidate for obtaining the adiabatic evolution of a resonant inspiral 
with a phase accuracy better than the order of unity.

The other point is that 
our new method does not reduce the importance of developing the 
direct self-force calculation, which has recently undergone 
rapid progress. 
We think that these two methods are complementary. Relatively high 
precision will be required in evaluating the effect that may cause
an orbital phase error enhanced by an inverse power of the mass ratio. 
Numerical evaluation of such quantities might be a tough problem 
if we cannot take advantage of the Teukolsky formalism that reduces 
the problem into one to solve one-dimensional radial functions. 
Even if achieving high accuracy numerically is not so difficult, 
our method will give reference numbers which are useful to check the 
accuracy of the results obtained by direct self-force calculations.

Before closing this paper, we would like to mention how to extend our 
method to the case of the gravitational radiation reaction. Since our 
method is based on the mode sum regularization in which each mode 
contribution is gauge invariant, we do not have to concern ourselves with the 
gauge adjustment between the symmetric part and the $S$-part. Even for 
the gravitational case, the contribution from the $S$-part evaluated 
in the Lorenz gauge is almost identical to the scalar case discussed 
in this paper. To calculate the contribution from the symmetric part, 
we can use the standard Teukolsky formalism with the aid of the 
established method of reconstructing the metric perturbation from the 
master variables
\cite{Chrzanowski:1975wv,Wald:1978vm,Kegeles:1979an,Ori:2002uv}.
The metric perturbation reconstructed in this manner is given in the 
so-called radiation gauge, in which a singularity runs from the 
particle's trajectory in the radial null direction. However, almost 
all regularized trajectories shifted in the Killing directions do not 
cross this singularity. Therefore we expect that the presence of this 
radial singularity will not cause any trouble. Although we need further 
study for practical implementation, we think that our method can be 
extended to the gravitational case, as briefly described above.

\acknowledgments
We thank  \'{E}anna \'{E}. Flanagan for his constructive comments on our work. 
S.I. is also grateful for his warm hospitality at Cornell 
University where the final part of this work was completed.
S.I. acknowledges the support of the Grant-in-Aid for JSPS Fellows, 
No. 24-4281.
H.N. and T.T. acknowledge the support of the Grand-in-Aid for 
Scientific Research (No. 24103006).
R.F. is grateful for the support of the European Union FEDER funds, 
the Spanish Ministry of Economy and Competitiveness (Project No. 
FPA2010-16495 and No. CSD2007-00042), and the Conselleria 
d'Economia Hisenda i Innovacio of the Govern de les Illes Balears. 
R.F. also appreciates the warm hospitality at the Yukawa Institute 
for Theoretical Physics, Kyoto University,  
where part of this work was completed.
T.T. is also supported by the Grand-in-Aid for 
Scientific Research 
(No. 21111006, No. 21244033, No.24103001).
We also received informative feedback on our work from the 
participants of the YITP Long-term Workshop YITP-T-12-03 on 
``Gravity and Cosmology 2012'' at the Yukawa Institute
for Theoretical Physics at Kyoto University.
Finally, this work was supported by the Grant-in-Aid for the Global COE Program
``The Next Generation of Physics, Spun from Universality and Emergence''
from the Ministry of Education, Culture, Sports, Science and Technology
of Japan. 
%

\appendix

\section{The possible pattern for the long-term orbital 
evolution crossing the resonance}
\label{app:evolution}
The final goal of this project is to understand the long-term orbital 
evolution crossing the resonance in the gravitational case. 
Toward this ambitious goal, 
this appendix is dedicated to quantitatively 
discussing how we can understand the long-term evolution 
of the resonant inspiral, as a future application of our formalism.
In this appendix, we consider the gravitational case: 
a point particle with the rest mass $\mu$ moves along 
the quasi-resonant orbits around a Kerr black hole with the rest mass 
$M$ and the spin angular momentum $aM$. 
To apply the following discussion to the scalar model, 
we just need to replace $\mu$ with $(q^2 / \mu)$ in the results, 
where $q$ is the scalar charge of a point scalar particle.

As was discussed in Refs.~\cite{Mino:2003yg,Tanaka:2005ue}, 
as far as we consider the first-order self-force in 
the osculating orbit approximation, 
the long-term orbital evolution is basically determined by 
tracing the evolution of slowly changing variables, \ie, 
the constants of motion $I^i$, and 
the orbital frequencies as their functions, $\Upsilon_a(I^i)$. 
We only need to know the non-oscillating part of the rates of change
of $I^i$ and $\Upsilon_a(I^i)$ for a generic 
orbit as the corrections due to the self-force.
In particular, we only need to know the former if it is allowable
to neglect $O((M/\mu)^0)$ error in the orbital phase.
This observation is essentially equivalent to the 
more systematic two-time scale analysis given in
Ref.~\cite{Hinderer:2008dm}.

Near the resonance, however, the adiabatic evolution of the orbit 
cannot be described by tracing the constants of motion 
$I^i$ because the self-force also depends on the offset phase $\Dl$, 
the relative phase between the $r$- and $\theta$ -oscillations, 
as discussed in the main text.
Neglecting the part that oscillates in the orbital time scale, 
the time derivatives of $I^i$ are determined as functions of 
$I^i$ and $\Dl$, which we denote by $\dot I^i(I^i,\Dl)$.
At the same time, however, $\Dl$ also evolves unless 
the exact resonance condition is satisfied, and the resonance
orbit effectively comes back to the off-resonance orbit 
when the offset phase starts to change rapidly.
We examine possible patterns of evolution of $\{I^i,\Dl \}$ 
below.

Suppose that the resonance condition is only approximately satisfied, 
but its deviation is small. 
In this situation, $\dot{I^i}$ may still depend on $\Dl$.
Moreover, we are also allowed to replace $I^i$ 
in $\dot{I^{i}} (I^{i}, \Dl)$ by a fixed value at
the exact resonance point, which we denote $I^{i}_{\rm res}$.
This can be understood in the following manner.
The typical time scale for crossing the resonance point is estimated as 
$\tau_{\rm res} \sim O(\Omega_{\phi} \sqrt{M /\mu})$. 
During the resonance crossing, 
the net change of the constants of motion
from $I^{i}_{\rm res}$ is estimated as 
$\Delta I^{i} \sim \dot{I^{i}} \tau_{\rm res} 
\sim O(\Omega_{\phi} \sqrt{\mu/M})$
\cite{Tanaka:2005ue, Hinderer:2008dm}.
This means that the correction to $I^{i}_{\rm res}$ is always 
suppressed by a factor of $O(\sqrt{\mu/M})$ in the resonance crossing.
Based on this observation, we can estimate the error in 
$\dot{I^{i}}$ that is associated with the replacement of 
its argument from $I^{i}$ to $I^{i}_{\rm res}$ as 
\begin{equation}\label{dI-dt}
\frac{d}{dt} \left( \log I^{i}  (I^{i}, \Dl)  \right) = 
\frac{d}{dt} \left( \log I^{i}_{\rm res}  (I^{i}, \Dl)  \right)
+ O\left(\Omega_{\phi} \left(\frac{\mu}{M}\right)^{3/2} \right).
\end{equation}
After passing the resonance, this error term causes the shifts 
of the constants of motion, which will be evaluated as 
$\Delta I^{i}_{\rm err}/ I^{i} \sim O(\mu/M)$. 
However, the net phase error due to these shifts, which 
is simply given by $\Delta I^{i}_{\rm err}$ times $\tau_{\rm rad}$,  
is at most $O((M/\mu)^0)$.
This phase error is small compared with what we are interested in here.
Thus, we conclude that 
the evolution of the constants of motion during the resonance crossing is 
given by 
$\dot{I^{i}} (I^{i}_{\rm res}, \Dl)$.

Next, we discuss the evolution of the offset phase, $\Dl$. In the
situation considered here, there is no longer an exact common resonance
period $\Lambda$. Instead, we introduce the approximate common period 
$\tilde{\Lambda}$ by 
\begin{equation}
\tilde{\Lambda} := j_\theta \Lambda_\theta =
\frac{2 \pi j_{\theta}}{\Upsilon_\theta}.
\end{equation}
During the period $\tilde{\Lambda}$, the offset phase is shifted
by
\begin{equation}
\delta(\Delta\lambda)
= \tilde{\Lambda} - j_r \Lambda_r
= \frac{\Lambda_r \Lambda_\theta}{2\pi} \Delta\Upsilon,
\end{equation}
where $\Delta\Upsilon:=j_\theta\Upsilon_r-j_r\Upsilon_\theta$.   
When $\Delta\Upsilon = 0$, we have an exact resonance.
In the adiabatic regime, we are also allowed to replace the time derivative of 
$\Delta \lambda$ with the averaged value over one period $\tilde{\Lambda}$.
As a result, 
the evolution of $\Delta\lambda$ is described as 
\begin{equation}\label{Deltalambdaevolv}
\left \langle
\frac{d \Delta \lambda}{dt}
\right \rangle_t
= \Upsilon_t^{-1} \frac{\delta(\Delta\lambda)}{\tilde{\Lambda}}
= \frac{\Delta\Upsilon}{j_\theta \Upsilon_t \Upsilon_r}.
\end{equation}

Now, we investigate the evolution of $\Delta\lambda$ around a
resonance point in more detail. 
For this purpose, focusing on the time derivative of 
$\langle d\Delta\lambda/dt \rangle_t$,
we consider the two cases separately, depending on whether or not
there exists $\Delta\lambda_c$ such that 
\begin{equation}
{d \over dt} \left \langle
\frac{d \Delta \lambda}{dt}
\right \rangle_t
=
\frac{\partial}{\partial I^j} \left(
\frac{\Delta\Upsilon}{j_\theta \Upsilon_t \Upsilon_r}
\right) \dot I^j(I^i,\Delta\lambda_c)
= 0
\end{equation}
(see also Ref.~\cite{Gair:2011mr} for the classification of
resonance orbits). 
The latter case has already been discussed in Ref.~\cite{Tanaka:2005ue} and 
is essentially the same as the transient resonance 
analyzed in Refs.~\cite{Flanagan:2010cd,Flanagan:2012kg}.
In this case, $\Delta\lambda$ varies with $O(1)$ on the time
scale of $O(\Omega_\phi / \sqrt{\mu})$ around $\Delta\Upsilon=0$,
during which $I^i$ evolves by $O(\sqrt{\mu})$ relatively.
The effect of the resonance alters the evolution of $I^i$
which leads to 
the phase corrections of $O(\sqrt{\mu} \Omega_\phi \Delta T )$, 
where $\Delta T$ is the time interval between the resonance 
and the plunge. As $\Delta T$ typically scales like 
$(\propto 1/\mu)$, 
the phase corrections due to this effect can be significantly 
large $(\propto 1/\sqrt{\mu})$ 
if the inspiral is in the extreme mass ratio regime, $M \gg \mu$.

A more interesting situation may happen if $\Delta\lambda_c$ exists. 
This case corresponds to the so-called sustained resonance
\cite{Flanagan:2010cd}.
When $\Delta\lambda_c$ is sufficiently close to $\Delta\lambda_c$, 
we obtain, with Taylor expansion around $\Delta\lambda_c$,
\begin{eqnarray}
{d \over dt} \left \langle
\frac{d \Delta \lambda}{dt}
\right \rangle_t
=
-\varpi^2(I^i) (\Delta\lambda-\Delta\lambda_c(I^i)), 
\end{eqnarray}
where
\[
 \varpi^2(I^i) := 
 - \frac{\partial}{\partial I^j} \left(
 \frac{\Delta\Upsilon}{j_\theta \Upsilon_t \Upsilon_r}
 \right)
 {\partial \dot I^j \over \partial\Delta\lambda}
 (I^i,\Delta\lambda_c(I^i)).
\]
Since the time scale $1/|\varpi|$  scales like
$(\propto 1/\sqrt{\mu})$, it is likely to be much shorter than the 
radiation reaction time scale, which scales like $(\propto 1/{\mu})$. 
If $\varpi^2>0$, the above system behaves like a harmonic 
oscillator with its potential minimum varying adiabatically. 
In this case, $\Delta\lambda$ oscillates around $\Delta\lambda_c$ 
with the amplitude $\propto 1/\sqrt{\varpi}$. 
This {evolution} may last as long as the condition $\varpi^2>0$ 
and the existence of $\Delta\lambda_c$ are maintained.  
Therefore, in principle, the resonance may last for 
the radiation reaction time scale.
This time scale is much longer than we naively expect when 
the possible presence of $\Delta\lambda_c$ is neglected. 
If $\varpi^2<0$, the value of $\Delta\lambda$ close to 
$\Delta\lambda_c$ is disfavored. The time scale for the variation 
of $\Delta\lambda$ is again 
$\propto 1/\sqrt{\mu}$ as in the case of the absence of $\Delta\lambda_c$. 
Hence, the corrections to the phase evolution are also of the same order 
as before.
Although the post-Newtonian analysis~\cite{Flanagan:2010cd} and the 
gravitational wave fluxes from a resonance orbit~\cite{Flanagan:2012kg} 
suggest that the sustained resonance may be absent in inspirals
around the  Kerr black hole, or may exist only in very restricted orbital
parameter regions even if it is achieved,  
the existence of the sustained resonance in the inspirals into 
the Kerr black hole is still an open question.

\section{Derivation of Eq.~(\ref{eq:dQdt-sym2})}
\label{app:simplify-dQdt-sym}
First, to show Eqs.~(\ref{eq:Gsym-proper1}) and
(\ref{eq:Gsym-proper2}), we clarify the properties of
$z^\mu(\lambda)$ and $G^{({\rm sym}-S)}(x,x')$, respectively.
For resonance orbits, $r(\lambda)$ and $\theta(\lambda)$ are
periodic with period $\Lambda(=j_r\Lambda_r=j_\theta\Lambda_\theta)$
as shown in Eq.~(\ref{eq:r-theta-motions}),
while $t(\lambda)$ and $\phi(\lambda)$ have secular evolution
terms proportional to $\lambda$ in addition to oscillatory parts
with period $\Lambda$ as Eqs.~(\ref{geodesic-t}) and
(\ref{phi-oscillation}). Taking account of these facts,
we find that, under the transformation
$\lambda \rightarrow \lambda+\Lambda$, $z^\mu(\lambda)$
transforms as
\begin{equation} \label{eq:orbit-displace}
z^\mu(\lambda+\Lambda) =
z^\mu(\lambda)
+ \left[\Upsilon_t \xi_{(t)}^\mu
+ \Upsilon_\phi \xi_{(\phi)}^\mu \right] \Lambda
=
z^\mu(\lambda)
+ \Upsilon_t \Lambda \xi^\mu(0),
\end{equation}
where $\xi^\mu(\zeta)$ is defined in Eq.~(\ref{helical}).
This means that the trajectory is displaced in a Killing
direction.

As for $G^{({\rm sym}-S)}(x,x')$, we find two properties.
Due to the stationarity and axisymmetry of the Kerr spacetime,
the Green functions are invariant under the displacement along
an arbitrary Killing vector field:
\begin{equation} \label{eq:G-displace}
G(x+A\xi(\zeta), x'+A\xi(\zeta))
=  G(x, x'),
\end{equation}
where $A$ is an arbitrary constant.
From Eqs. (\ref{eq:orbit-displace}) and (\ref{eq:G-displace}),
we can derive Eq.~(\ref{eq:Gsym-proper1}) as
\begin{eqnarray*}
G(z(\lambda+\Lambda), z(\lambda'+\Lambda))
&=&
G(z(\lambda)+\Upsilon_t\Lambda\xi(0),
z(\lambda')+\Upsilon_t\Lambda\xi(0)) \\
&=& G(z(\lambda), z(\lambda')).
\end{eqnarray*}
The relation (\ref{eq:G-displace}) simply tells that the
Green functions depend on $t$ and $t'$ through $t-t'$, and
on $\phi$ and $\phi'$ through $\phi-\phi'$, respectively. 
Based on this understanding, 
we find that Eq.~(\ref{eq:Gsym-proper2}) holds.

Next, we show that the first term on the right-hand side
of Eq.~(\ref{eq:dQdt-sym1}) vanishes
by using Eqs.~(\ref{eq:Gsym-proper0}),
(\ref{eq:Gsym-proper1}), and (\ref{eq:Gsym-proper2}).
We focus only on the $t$-derivative part, but one can
show that the $\phi$-derivative part vanishes in the
same way.
By using the property of Eq.(\ref{eq:Gsym-proper1}), we can
replace the long-time average over $\lambda$ with the average
over one period $\Lambda$ as
\begin{eqnarray}
&& \hspace*{-1cm}
\int_{-\infty}^{\infty} d\lambda'
\Sigma[z(\lambda')] \left\langle
\left[ \Sigma(x) \partial_t G(x,z(\lambda'))
\right]_{x=z(\lambda)}
\right\rangle_\lambda
\nonumber \\ &=&
\lim_{T\rightarrow\infty}\frac{1}{2T}\int_{-T}^{T} d\lambda
\int_{-\infty}^{\infty} d\lambda'
\Sigma[z(\lambda)] \Sigma[z(\lambda')]
\left[ \partial_t G(x,x')
\right]_{x=z(\lambda), x'=z(\lambda')}
\nonumber \\ &=&
\lim_{N\rightarrow\infty}\frac{1}{2N\Lambda}
\sum_{n=-N}^{N-1}\int_{n\Lambda}^{(n+1)\Lambda}
\!\!\!\! d\lambda
\int_{-\infty}^{\infty} d\lambda'
\Sigma[z(\lambda)] \Sigma[z(\lambda')]
\left[ \partial_{t} G(x,x')
\right]_{x=z(\lambda), x'=z(\lambda')}
\nonumber \\ &=&
\lim_{N\rightarrow\infty}\frac{1}{2N\Lambda}
\sum_{n=-N}^{N-1}\int_{0}^{\Lambda} d\lambda
\int_{-\infty}^{\infty} d\lambda'
\Sigma[z(\lambda)] \Sigma[z(\lambda')]
\left[ \partial_{t} G(x,x')
\right]_{x=z(\lambda), x'=z(\lambda')}
\nonumber \\ &=&
\frac{1}{\Lambda}\int_{0}^{\Lambda} d\lambda
\int_{-\infty}^{\infty} d\lambda'
\Sigma[z(\lambda)] \Sigma[z(\lambda')]
\left[ \partial_{t} G(x,x')
\right]_{x=z(\lambda), x'=z(\lambda')}.
\label{eq:partial_t_part1}
\end{eqnarray}
In the third equality, we transformed $(\lambda, \lambda')$
to $(\lambda-n\Lambda, \lambda'-n\Lambda)$, and used
Eq.~(\ref{eq:Gsym-proper1}) and
$\Sigma[z(\lambda-n\Lambda)]=\Sigma[z(\lambda)]$
(note that $\Sigma(x)$ is a function of $r$ and $\theta$).
We also rewrite the integrals over $\lambda$ and $\lambda'$ as
\begin{eqnarray}
&& \hspace*{-1cm}
\frac{1}{\Lambda}\int_{0}^{\Lambda} d\lambda
\int_{-\infty}^{\infty} d\lambda'
\Sigma[z(\lambda)] \Sigma[z(\lambda')]
\left[ \partial_{t} G(x,x')
\right]_{x=z(\lambda), x'=z(\lambda')}
\nonumber \\ &=&
\frac{1}{\Lambda}
\sum_{n=-\infty}^\infty
\int_{0}^{\Lambda} d\lambda
\int_{n\Lambda}^{(n+1)\Lambda} \!\!\!\! d\lambda'
\Sigma[z(\lambda)] \Sigma[z(\lambda')]
\left[ \partial_{t} G(x,x')
\right]_{x=z(\lambda), x'=z(\lambda')}
\nonumber \\ &=&
\frac{1}{\Lambda}
\sum_{n=-\infty}^\infty
\int_{-n\Lambda}^{-(n-1)\Lambda}
\!\!\!\! d\lambda
\int_{0}^{\Lambda} d\lambda'
\Sigma[z(\lambda)] \Sigma[z(\lambda')]
\left[ \partial_{t} G(x,x')
\right]_{x=z(\lambda), x'=z(\lambda')}
\nonumber \\ &=&
\frac{1}{\Lambda}
\int_{-\infty}^{\infty} d\lambda
\int_{0}^{\Lambda} d\lambda'
\Sigma[z(\lambda)] \Sigma[z(\lambda')]
\left[ \partial_{t} G(x,x')
\right]_{x=z(\lambda), x'=z(\lambda')}.
\label{eq:partial_t_part2}
\end{eqnarray}
Further, for the Green function $G^{({\rm sym}-S)}(x,x')$,
by exchanging the labels in the last line of
Eq.~(\ref{eq:partial_t_part2})
as $\lambda\leftrightarrow\lambda'$
and $x \leftrightarrow x'$, we obtain
\begin{eqnarray}
&& \hspace{-1cm}
\frac{1}{\Lambda}
\int_{-\infty}^{\infty} d\lambda
\int_{0}^{\Lambda} d\lambda'
\Sigma[z(\lambda)] \Sigma[z(\lambda')]
\left[ \partial_{t} G^{({\rm sym}-S)}(x,x')
\right]_{x=z(\lambda), x'=z(\lambda')}
\nonumber \\ &=&
\frac{1}{\Lambda}
\int_{0}^{\Lambda} d\lambda
\int_{-\infty}^{\infty} d\lambda'
\Sigma[z(\lambda)] \Sigma[z(\lambda')]
\left[ \partial_{t'} G^{({\rm sym}-S)}(x',x)
\right]_{x=z(\lambda), x'=z(\lambda')}
\nonumber \\ &=&
\frac{1}{\Lambda}
\int_{0}^{\Lambda} d\lambda
\int_{-\infty}^{\infty} d\lambda'
\Sigma[z(\lambda)] \Sigma[z(\lambda')]
\left[ - \partial_{t} G^{({\rm sym}-S)}(x,x')
\right]_{x=z(\lambda), x'=z(\lambda')},
\end{eqnarray}
where we used Eqs.~(\ref{eq:Gsym-proper0}) and
(\ref{eq:Gsym-proper2}) in the last equality.
Therefore, we find
\begin{equation} \label{eq:partial_t_part}
\int_{-\infty}^{\infty} d\lambda'
\Sigma[z(\lambda')] \left\langle
\left[ \Sigma(x) \partial_t G^{({\rm sym}-S)}(x,z(\lambda'))
\right]_{x=z(\lambda)}
\right\rangle_\lambda
= 0.
\end{equation}

As for the $\Delta\lambda$-derivative part, we transform
\begin{eqnarray}
&& \hspace{-1cm}
\int_{-\infty}^\infty d\lambda' \left[
\Sigma(x') \frac{d}{d(\Delta\lambda)}
\left\langle \Sigma[z(\lambda)]
G^{({\rm sym}-S)}(z(\lambda), x')
\right\rangle_\lambda
\right]_{x'=z(\lambda')}
\nonumber \\ &=&
\frac{1}{\Lambda}
\int_0^\Lambda d\lambda
\int_{-\infty}^\infty d\lambda'
\left[ \Sigma(x)
\frac{d}{d(\Delta\lambda)}
\left\{ \Sigma[z(\lambda')]
G^{({\rm sym}-S)}(x, z(\lambda'))
\right\} \right]_{x=z(\lambda)},
\label{eq:partial_dl_part1}
\end{eqnarray}
where we replaced the long-time average with respect to
$\lambda$ with the average over one orbital period $\Lambda$
as shown in Eq.~(\ref{eq:partial_t_part1}),
and exchanged the labels as
$\lambda\leftrightarrow\lambda'$ and $x \leftrightarrow x'$,
and the domains of integration with respect
to $\lambda$ and $\lambda'$ in the same way as in
Eq.~(\ref{eq:partial_t_part2}).
Using Eq.~(\ref{eq:partial_dl_part1}), we find
\begin{eqnarray}
&& \hspace{-1cm}
\int_{-\infty}^\infty d\lambda' \left[
\Sigma(x') \frac{d}{d(\Delta\lambda)}
\left\langle \Sigma[z(\lambda)]
G^{({\rm sym}-S)}(z(\lambda), x')
\right\rangle_\lambda
\right]_{x'=z(\lambda')}
\nonumber \\ &=&
\frac{1}{2\Lambda}
\int_0^\Lambda d\lambda
\int_{-\infty}^\infty d\lambda'
\left\{
\left[ \Sigma(x')
\frac{d}{d(\Delta\lambda)}
\left\{ \Sigma[z(\lambda)]
G^{({\rm sym}-S)}(z(\lambda), x')
\right\} \right]_{x'=z(\lambda')}
\right. \cr && \left. \hspace{3cm} +
\left[ \Sigma(x)
\frac{d}{d(\Delta\lambda)}
\left\{ \Sigma[z(\lambda')]
G^{({\rm sym}-S)}(x, z(\lambda'))
\right\} \right]_{x=z(\lambda)}
\right\}
\nonumber \\ &=&
\frac{1}{2\Lambda}
\int_0^\Lambda d\lambda
\int_{-\infty}^\infty d\lambda'
\frac{d}{d(\Delta\lambda)}
\left[ \Sigma[z(\lambda)] \Sigma[z(\lambda')]
G^{({\rm sym}-S)}(z(\lambda), z(\lambda')) \right]
\nonumber \\ &=&
\frac{1}{2} \frac{d}{d(\Delta\lambda)}
\int_{-\infty}^\infty d\lambda'
\left\langle \Sigma[z(\lambda)] \Sigma[z(\lambda')]
G^{({\rm sym}-S)}(z(\lambda), z(\lambda'))
\right\rangle_{\lambda'}.
\label{eq:partial_dl_part}
\end{eqnarray}
Removing the $t$- and $\phi$-derivative parts from
Eq.~(\ref{eq:dQdt-sym1}) and using
Eq.~(\ref{eq:partial_dl_part}),
we finally obtain Eq.~(\ref{eq:dQdt-sym2}).



\end{document}